\documentclass[iop,twocolappendix,numberedappendix]{emulateapj}
\usepackage{graphicx}
\usepackage{subfigure}
\usepackage{multirow}
\usepackage{amssymb}
\usepackage{natbib}
\usepackage{rotating}
\usepackage[bookmarks=false, colorlinks=true, citecolor=blue, backref=true]{hyperref}
\usepackage{apjfonts}

\shorttitle{Galaxy Pairs in COSMOS -- Merger Rate Evolution Since $z= 1$}
\shortauthors{C.K. Xu et al.}

\newcommand{\lsim}{\, \lower2truept\hbox{${< \atop\hbox{\raise4truept\hbox{$\sim$}}}$}\,}
\newcommand{\gsim}{\, \lower2truept\hbox{${> \atop\hbox{\raise4truept\hbox{$\sim$}}}$}\,}

\newcommand{\oneskip}{\vskip\baselineskip}

\begin{document}

\slugcomment{{\bf Draft 10}; \today}

\title{Major-Merger Galaxy Pairs in the COSMOS Field --- \break
       Mass Dependent Merger Rate Evolution Since $z = 1$}

\author{C. Kevin Xu \altaffilmark{1}, 
Yinghe Zhao\altaffilmark{2},  
N. Scoville\altaffilmark{3},
P. Capak\altaffilmark{4},
N. Drory\altaffilmark{5,6},
Y. Gao\altaffilmark{2}
}
\altaffiltext{1}{Infrared Processing and Analysis Center, 
California Institute of Technology 100-22, Pasadena, CA 91125, USA}
\altaffiltext{2}{Purple Mountain Observatory, Chinese Academy of Sciences, 2 West Beijing Road, Nanjing 210008, China}
\altaffiltext{3}{California Institute of Technology, MC 105-24, 
1200 East California Boulevard, Pasadena, CA 91125, USA}
\altaffiltext{4}{Spitzer Science Center, California Institute of Technology, Mail Stop 220-6, Pasadena, CA 91125, USA}
\altaffiltext{5}{Instituto de Astronom\'ia, Universidad Nacional Aut\'onoma de M\'exico, A.P.\ 70-264, 04510 M\'exico, D.F., M\'exico}
\altaffiltext{6}{Max-Planck Institut f\"ur extraterrestrische Physik, 
Giessenbachstrasse, 85748 Garching, Germany}

\accepted{Nov.~22, 2011}

\begin{abstract}
We present results of a statistical study of the cosmic evolution of
the mass dependent major-merger rate since $\rm z=1$. A stellar mass
limited sample of close major-merger pairs (the CPAIR sample) was selected from
the archive of the COSMOS survey. Pair fractions at different
redshifts derived using the CPAIR sample and a local K-band selected
pair sample show no significant variations with stellar mass. The pair
fraction exhibits moderately strong cosmic evolution, with the
best-fitting function of $\rm f_{pair}=10^{-1.88(\pm 0.03)}
(1+z)^{2.2(\pm 0.2)}$. The best-fitting function for the merger rate
is $\rm R_{mg}\, (Gyr^{-1})=0.053\times (M_{star}/10^{10.7}
M_{\sun})^{0.3}(1+z)^{2.2}/(1+z/8)$. This rate implies that galaxies
of $\rm M_{star} \sim 10^{10\hbox{--}11.5}\; M_{\sun}$ have undergone
$\sim 0.5$ -- 1.5 major-mergers since z=1. Our results show that, for
massive galaxies ($\rm M_{star}\geq 10^{10.5}\; M_\sun$) at $z\leq 1$,
major mergers involving star forming galaxies (i.e. wet and mixed
mergers) can account for the formation of both ellipticals and red
quiescent galaxies (RQGs). On the other hand, major mergers cannot be
responsible for the formation of most low mass ellipticals and RQGs of
$\rm M_{star}\lsim 10^{10.3}\; M_\sun$. Our quantitative estimates
indicate that major mergers have significant impact on the stellar mass
assembly of the most massive galaxies ($\rm M_{star}\geq 10^{11.3}\;
M_\sun$), but for less massive galaxies the stellar mass assembly is
dominated by the star formation. Comparison with the mass dependent
(U)LIRG rates suggests that the frequency of major-merger events is
comparable to or higher than that of (U)LIRGs.
\end{abstract}


\section{Introduction}
Galaxy mergers have been fascinating astronomers for a long time, ever
since they were recognized (see the
review of \citet{Schweizer1996}). Major mergers of
galaxies of nearly equal mass stand out because of the more spectacular
tidal and dynamical 
effects \citep{Toomre1978}, and many nearby major
mergers have been extensively studied \citep{Toomre1972, Whitmore1995,
Hibbard1996, Hibbard1999, Xu2000, Wang2004}. 
It has been well documented that major mergers
can induce enhanced star formation \citep{Kennicutt1987, Xu1991}, 
trigger extreme starbursts and active galactic nuclear (AGN)
activities \citep{Sanders1988, Sanders1996, Dasyra2006}, 
and transform spiral galaxies into elliptical galaxies
\citep{Toomre1978, Schweizer1982, Genzel2001}. They dominate among
the extreme starbursts such as luminous infrared galaxies (LIRGs, with
SFR $\rm \gsim 20\; M_\sun$ yr$^{-1}$) and ultra-luminous infrared
galaxies (ULIRGs, with SFR $\rm \gsim 200\; M_\sun$ yr$^{-1}$; 
\citealt{Sanders1996}).
On the other hand, statistically, major mergers play
minor roles in the processes such as star formation and mass growth of
$\rm z \sim 0$ galaxies in general. Only $\sim$ 1 -- 2 percent of
galaxies are involved in close major mergers \citep{Xu2004, Patton2008,
Domingue2009}, and only $\sim$ 2 -- 3 percents of star formation
rate density (SFRD) in the z=0 universe is due to close major mergers 
\citep{Xu2010}.

Are mergers more important in the earlier universe? 
Indeed, in the hierarchical
structure formation paradigm of the contemporary cosmology, galaxy and
dark matter halo (DMH) merging is one of the most significant
processes affecting the evolution of structures in the early universe,
and is largely responsible for the growth of massive dark matter halos and
the buildup of galaxies \citep{Kauffmann1993, Lacey1993,
Khochfar2005}.  Many observations of intermediate/high redshift
peculiar galaxies and galaxy pairs have found strong evolution in the
merger rate, up to $\rm (1+z)^{{\hbox {3-6}}}$ \citep{Brinchmann1998,
LeFevre2000, Conselice2003, Bell2006, 
Conselice2006, Conselice2009, Kampczyk2007,
Kartaltepe2007, Rawat2008}, and show evidence for mergers to dominate
the total star formation rate in the universe of $z \gsim 1$ \citep{Zheng2004,
Hammer2005, Bridge2007}.  On the other hand,
weak merger rate evolution ($\rm \sim (1+z)^{0.5}$) has been found by
other studies of intermediate/high redshift mergers \citep{Carlberg2000,
Lin2004, Lotz2008, Robaina2010, Man2011}, and many
authors have argued that at $\rm z\gsim 1$ the SFRD 
is still predominantly contributed by isolated late
type galaxies rather than by mergers \citep{Flores1999, Bell2005, 
Lotz2008, Jogee2009}.
Furthermore, mergers may no longer be the predominant population
among LIRGs and even ULIRGs at $\rm z \gsim 1$\citep{Melbourne2005, Daddi2010}.

In this paper, we address the controversy on the merger rate
evolution using data in the COSMOS field \citep{Scoville2007}.
Major reasons for the controversy include different definitions of
major mergers, errors in the merger time scale, and biases in the
merger sample selections (see \citealt{Xu2010, Hopkins2010b}). 
Studies using merger
samples selected from peculiar galaxies 
\citep{LeFevre2000, Conselice2003, Conselice2006, Jogee2009, Conselice2009} 
are vulnerable to contaminations
due to minor mergers and to uncertainties of the time scale for the
detectable peculiarity  such as
tidal tails, bridges, plumes, and other distortions
\citep{Lotz2010}. In contrast, we selected our merger
samples from close ($\rm 5 \leq r_{proj} \leq 20\; h^{-1}\; kpc$) major-merger
pairs (stellar mass ratios $\leq 2.5$). These pairs
have reasonably well understood 
merger time scales \citep{Kitzbichler2008, Lotz2010}. 

Our pair sample is selected from the photo-z catalog of the
COSMOS field \citep{Ilbert2009} and is stellar mass limited,
including massive galaxies ($\rm M_{star} \geq 10^{9} M_{\sun}$)
in the photo-z range of $\rm 0.2 \leq z_{phot} \leq 1$.  The COSMOS sample
has the best photo-z's, measured using
data of $\sim 30$ photometric bands covering the entire UV --
infrared range, for more than 100,000 galaxies with nearly 100 percent
completeness \citep{Ilbert2009}.  This enables us to obtain a pair
sample that is $\sim 70\%$ complete.  By comparison the pair samples
in the studies of \citet{Patton2002}, \citet{Lin2004}, 
and \citet{Bundy2009}, using pairs selected from spectroscopic
surveys, are only $\sim 10$ -- 20\% complete. Given the rather complex
spectroscopic selection functions in those studies, the corrections
for the incompleteness may lead to substantial uncertainties in the
results. 

Photo-z selected pairs of $\rm 0.2 \leq z_{phot} \leq 1.2$ in the COSMOS
field were studied by \citet[hereafter K07]{Kartaltepe2007}. Their
sample is different from ours in two major respects: (1) It is not
confined to major mergers (i.e. no constraint on the mass ratios or
luminosity ratios between primaries and secondaries); (2) it is a
absolute magnitude limited sample ($\rm M_{V} \leq -19.8$).  
With more rigorously
defined major-merger pair samples, our study shall improve
upon the results of K07. Also, with well determined stellar mass for
every galaxy in the sample, 
we shall study the mass dependence of the merger rate evolution.

We will confine our analysis to galaxy pairs of $\rm z \leq 1$,
because the photo-z's and stellar mass estimates of
$z > 1$ galaxies are less accurate (\citealt{Ilbert2010}).  We will derive
the merger rates for galaxies of different stellar masses and
redshifts, and separate dry mergers (E+E pairs) and wet/mixed mergers
(S+S and S+E mergers). The selection of
COSMOS pairs in the redshift range of $\rm 0.2 \leq z \leq 1$ is described
in Section 2.  Corrections for the incompleteness and for contaminations by
spurious pairs are presented  in Section 3. A z=0 pair
sample, which sets the local benchmark for the evolution study, is
presented  in Section 4.  The mass dependent merger rate and its
evolution since z=1 are presented in Section 5. The contributions of
major mergers to the galaxy assembly and elliptical galaxy formation
since z=1 is investigated in Section 6, and comparisons to the mass
and redshift dependence of the (U)LIRGs' abundance is in Section
7. Section 8 is devoted to a summary of our main results.  Throughout
this paper, we adopt the $\Lambda$-cosmology with $\rm \Omega_m=0.3$
and $\rm \Omega_\Lambda = 0.7$, and $\rm H_0= 70\; (km~sec^{-1}
Mpc^{-1})$.
\begin{deluxetable}{ccccccccccccc}
\tabletypesize{\normalsize}
\setlength{\tabcolsep}{0.05in} 
\tablenum{1}
\tablecaption{Parent Sample\label{tbl:parent}}
\tablehead{
 \colhead{$\rm z_{min}$} 
& &  \colhead{$\rm z_{max}$} 
& &  \colhead{Volume} 
& &  \colhead{$\rm \log(M_{min})$} 
& &  \multicolumn{5}{c}{Number of Galaxies$^\dagger$} 
\\
\cline{9-13} 
\\
& &
& & \colhead{($\rm 10^6 Mpc^3$)} 
& & \colhead{($\rm M_\sun$)} 
& & \colhead{$\rm SFGs$} 
& & \colhead{$\rm RQGs$} 
& & \colhead{$\rm Total$} 
}
\startdata
0.2 & & 0.4 & & 0.56 & &  9.0 & & 6787& &2039  & &8826 \\ 
0.4 & & 0.6 & & 1.23 & &  9.4 & & 6169& &1526  & &7695 \\ 
0.6 & & 0.8 & & 1.92 & &  9.8 & & 6745& &1981  & &8726 \\ 
0.8 & & 1.0 & & 2.53 & & 10.2 & & 6610& &2287  & &8897 \\ 
\hline
\enddata
\tablecomments{
\begin{description}
\item{$\dagger$} Number of galaxies with $\rm \log(M_{star}) \geq \log(M_{min})$.
\end{description}
}
\end{deluxetable}

\section{The COSMOS Pair (CPAIR) Sample}

We selected major-merger pair candidates using a parent sample of
galaxies constructed from that used by \citet[hereafter
  D09]{Drory2009} in their study of galaxy stellar mass function
(GSMF), which is in turn selected from the COSMOS photo-z catalog
\citep{Ilbert2009} using the following criteria: $\rm 0.2 \leq
z \leq 1$, $\rm K_s < 24$ and $i^+_{\rm AB} < 25.5$.

The D09 sample has 138001
galaxies, divided into four photo-z bins of width of $\Delta z= 0.2$,
and into ``active'' (star forming galaxies, SFGs) and ``passive'' 
(red quiescent galaxies, RQGs) populations according to
the SED type of the best fitting template  \citep{Ilbert2009}.
The stellar mass of galaxies, $\rm M_{star}$, is derived through a
stellar population synthesis model fitting (the Chabrier IMF),
using the photo-z and photometric data in the $u^{*}$ (CFHT),
$B_J$, $V_J$, g$^+$, r$^+$, i$^+$, z$^+$ (Subaru), $J$ (UKIRT) and 
$K_s$ (CFHT) bands. Typical uncertainties in $\rm M_{star}$ is between
0.1 dex and 0.3 dex at 68\% confidence level, depending on spectral
type and the S/N of the photometry (D09). 
In the four photo-z bins, the completeness limits for SFGs
and RQGs are $\rm \log(M_{star}/M_\sun) = [8.3, 8.9, 9.2, 9.4]$ and
$\rm \log(M_{star}/M_\sun) = [8.9, 9.2, 9.8, 10.1]$, respectively.

For galaxies in our parent sample, the stellar mass is taken
from D09. In order to be complete for both SFGs and RQGs, we imposed
a mass limit ($\rm \log(M_{min})$) on each of the
photo-z bins in the parent sample.
The mass limit and the number of galaxies above the limit
are listed in Table~\ref{tbl:parent}. 
There are 34144 galaxies in the parent sample.

The pair sample is also divided into four photo-z bins.
The selection criteria are:
\begin{description}
\item{(1)} The primary galaxy has $\rm \log(M_{star}) \geq \log(M_{lim})$,
with $\rm \log (M_{lim}/M_{\sun}) = [9.4, 9.8, 10.2, 10.6]$ for 
the four redshift bins, respectively. $\rm M_{lim}$'s are 
0.4 dex above the $\rm M_{min}$'s of the parent sample (Table~\ref{tbl:parent}).
\item{(2)} the
difference in $\rm M_{star}$
between the primary galaxy and the secondary galaxy is less than 0.4 dex:
$\rm \Delta log(M_{star}) \leq 0.4$.
\item{(3)} the redshift difference between the two components,
$\rm \Delta z_{phot} =  |z_{phot}^{pri} - z_{phot}^{2nd}|$, satisfies
$\rm \Delta z_{phot}/(1+z_{phot}^{pri}) \leq 0.03$.
\item{(4)} the projected physical separation ($\rm r_{proj}$) is in the range of 
$\rm 5 \leq r_{proj} \leq 20\; h^{-1}\; kpc$.
\end{description}

\setcounter{table}{1}
\begin{deluxetable*}{ccccccccccccccccccccc}
\tabletypesize{\normalsize}
\setlength{\tabcolsep}{0.05in} 
\tablenum{2}
\tablewidth{0pt}
\tablecaption{Sample of Paired Galaxies in COSMOS\label{tbl:sample}}
\tablehead{
 \colhead{$\rm z_{min}$} 
& &  \colhead{$\rm z_{max}$} 
& &  \multicolumn{5}{c}{Number of Galaxies$^\dagger$} 
& &  \colhead{$\rm \log(M_{lim})$} 
& &  \multicolumn{9}{c}{Number of Galaxies with $\rm M_{star} \geq M_{lim}$$^\ddagger$} 
\\
\cline{5-9} 
\cline{13-21} 
& &
& & \colhead{in iso.} 
& & \colhead{in multi.} 
& & \colhead{Total} 
& & \colhead{($\rm M_\sun$)} 
& & \colhead{in iso.} 
& & \colhead{in multi.} 
& & \colhead{Total} 
& & \colhead{SFGs$^*$} 
& & \colhead{RQGs$^*$} 
\\
& &
& & \colhead{pairs} 
& & \colhead{systems} 
& & 
& & 
& & \colhead{pairs} 
& & \colhead{systems} 
& & 
& &
& &
}
\startdata
0.2 & & 0.4 &&144 & &  6 & & 150 & &  9.4 & &128 & & 5 & & {\bf 133} & & 78& & 55 \\ 
0.4 & & 0.6 &&100 & &  3 & & 103 & &  9.8 & & 93 & & 3 & & {\bf  96} & & 61& & 35 \\ 
0.6 & & 0.8 &&144 & & 22 & & 166 & & 10.2 & &126 & &20 & & {\bf 146} & &109& & 37 \\ 
0.8 & & 1.0 &&174 & & 24 & & 198 & & 10.6 & &131 & &21 & & {\bf 152} & & 81& & 71 \\ 
\hline
\enddata
\tablecomments{
\begin{description}
\item{$\dagger$} Including all primaries and all secondaries.
\item{$\ddagger$} Including all primaries and those secondaries with
$\rm \log(M_{star}) \geq \log(M_{lim})$.
\item{$*$} SFGs (``active galaxies'') 
and RQGs (``passive galaxies'') classifications were taken from D09.
\end{description}
}
\end{deluxetable*}
\oneskip

Compared to the selection criteria for local pairs described in  
\citet{Xu2004},
we replaced the rest-frame K-band selection by a
stellar mass selection in criteria (1) and (2). 
The Spizer-IRAC 3.6$\mu m$ and 4.5$\mu m$ bands, which encompass
the rest-frame K-band emission for galaxies of $\rm 0.6 \lsim z \leq
1$, have relatively low angular resolution compared to the HST/ground-based
optical and NIR data. Using IRAC data would have
resulted in larger confusion errors in the stellar mass for pairs with
separation $\lsim 2''$. At the same time, it was shown in D09 
that for field galaxies of $\rm z \leq 1$ the stellar mass derived
using the HST/ground-based optical and NIR data are nearly identical
to those derived using data including the IRAC fluxes \citep{Ilbert2010}.

Criteria (1) and (2) guarantee that our pair sample is
not affected by the ``missing secondary'' bias (Xu et al. 2010).
Criterion (3) is set to minimize the incompleteness due to
the photo-z error while ensure that the reliability of the sample is
not significantly compromised. This issue will be addressed in detail in
Section 3.

Using these criteria, 417 pair candidates were selected. 
In order to exclude spurious
pairs due to imaging artifacts, visual inspections were carried out on
the HST-ACS images (F814 band, \citealt{Koekemoer2007}). Among 417
pair candidates, 335 were covered by the HST survey.  All pair
candidates outside the area of HST-ACS imaging (82/417 = 19.7\%) were
dropped from the final pair sample.    Fourteen
spurious pairs were identified: 8 have wrong astrometry (i.e. no
source appears at the sky coordinates) for at least one of the two
galaxies and 6 are pieces of single large disc galaxies.

Some galaxies were 
found repetitively in multiple pair candidates.
These 40 pair candidates consist of 17 triplets 
and 1 quartet, including 55 galaxies. The remaining
281 candidates are isolated pairs, including 562 galaxies. 
Because galaxies in triplets/quartets are less than 10\% of
the total sample, we will not distinguish them from paired
galaxies. Our final COSMOS Pair sample (hereafter CPAIR) 
includes {\bf 617} paired galaxies, found in both isolated pairs
and multiple systems. Among them, {\bf 527} (including both primaries and
secondaries) have $\rm \log(M_{star}) 
\geq \log(M_{lim})$, and the remaining (90, all secondaries) have
$\rm (\log(M_{lim})-0.4) \leq \log(M_{star}) < \log(M_{lim})$.
 Statistics of the sample are listed in Table~\ref{tbl:sample}.

\section{CPAIR Sample: Incompleteness and Spurious Pairs Fraction}
Much of the discrepancies between different results on merger rate evolution
can be attributed to various biases 
causing incompleteness (missing of true mergers) and contaminations of
spurious mergers in merger samples. Therefore it is
important to investigate thoroughly all such biases and correct them
in merger statistics.

\subsection{Incompleteness due to Missing Very Close Pairs}
Photometric data of the photo-z catalog \citep{Ilbert2009}
were obtained using the SExtractor in dual mode 
\citep{Bertin1996}. Images in all bands were degraded to a
common PSF of $\rm FWHM=1.5''$, 
and the photometry was done with a constant aperture
of $\rm r=1.5''$ \citep{Capak2007}. 
Because of the limited angular resolution of the photo-z catalog, very
close pairs with angular sepatation $\epsilon \lsim 2''$ are incomplete
in the pair sample. Exploiting the COSMOS HST-ACS lensing catalog
\citep{Leauthaud2007, Leauthaud2010}, we estimated this incompleteness
to be [0.01, 0.06, 0.08, 0.20] for the four redshift bins, with 
no significant mass dependence. 
The full analysis can be found in Appendix A.

\subsection{Incompleteness due to Photo-z Errors
and Spurious Pairs due to Projection} 
For pairs of $\epsilon > 2.0''$, major cause of the incompleteness
is due to photo-z errors, which have a non-negligible probability of
being so large that a real pair with $\rm \Delta v < 500\; km\;
sec^{-1}$ (corresponding to 
$\rm \Delta z/(1+z) < 0.0017$) can have a measured $\rm
\Delta z_{phot}/(1+z_{phot}) > 0.03$ and therefore be missed by the
CPAIR sample. Also, the photo-z selection criterion and the photo-z errors 
can introduce spurious pairs whose velocity difference $\rm
\Delta v$ is larger than $\rm 500\; km\; sec^{-1}$.  
Using Monte Carlo simulations, we estimated the incompleteness and the 
spurious pair fraction (hereafter SPF) to be
[0.21, 0.21, 0.23, 0.25] and [0.07, 0.08, 0.10, 0.09], repectively,
for the four refshift bins. The full analysis is presented in Appendix B.

\subsection{Clustering Effect on Spurious Pair Contaminations}

\setcounter{table}{2}
\begin{deluxetable*}{ccccccccccccccccc}
\tabletypesize{\normalsize}
\setlength{\tabcolsep}{0.05in} 
\tablenum{3}
\tablewidth{0pt}
\tablecaption{Completeness and Reliability Corrections for CPAIR Sample
\label{tbl:correction}}
\tablehead{
 \colhead{$\rm z_{min}$} 
& &  \colhead{$\rm z_{max}$} 
& &  \multicolumn{5}{c}{Completeness Correction} 
& &  \multicolumn{7}{c}{Reliability Correction}
\\
\cline{5-9} 
\cline{11-17} 
& &
& & \colhead{due to missing} 
& & \colhead{due to photo-z} 
& & \colhead{Combined} 
& & \colhead{due to random} 
& & \colhead{due to} 
& & \colhead{due to pairs of} 
& & \colhead{Combined} 
\\
& &
& & \colhead{very close pairs} 
& & \colhead{errors} 
& & 
& & \colhead{projection} 
& & \colhead{clustering} 
& & \colhead{$\rm \Delta v > 500km/sec$} 
& & \colhead{} 
}
\startdata
0.2 & & 0.4 &&$0.99\pm 0.01$ & &$0.79\pm 0.05$ & &$0.78\pm 0.05$ & &$0.93\pm 0.03$ & &$0.94\pm 0.05$&&$0.91\pm 0.03$ & &$0.80\pm 0.06$   \\ 
0.4 & & 0.6 &&$0.94\pm 0.03$ & &$0.79\pm 0.05$ & &$0.74\pm 0.05$ & &$0.92\pm 0.03$ & &$0.94\pm 0.05$&&$0.91\pm 0.03$ & &$0.79\pm 0.06$   \\ 
0.6 & & 0.8 &&$0.92\pm 0.03$ & &$0.77\pm 0.05$ & &$0.71\pm 0.05$ & &$0.90\pm 0.03$ & &$0.94\pm 0.05$&&$0.91\pm 0.03$ & &$0.77\pm 0.06$   \\ 
0.8 & & 1.0 &&$0.80\pm 0.04$ & &$0.75\pm 0.05$ & &$0.60\pm 0.05$ & &$0.91\pm 0.03$ & &$0.94\pm 0.05$&&$0.91\pm 0.03$ & &$0.78\pm 0.06$   
\enddata
\end{deluxetable*}
\oneskip

\citet{Bell2006} found that in the COMBO-17 survey, 
the projected two-point correlation functions of
massive galaxies  with $\rm 0.4 < z_{phot} \leq 0.8$ can be well fitted by
a power-law $\rm w(r)\propto r^{-\gamma}$ down to $\rm r = 15\; kpc$,
with the value of the power-index $\rm \gamma$ consistent with 2.
Based on this result (see also \citealt{Robaina2010}), 
we made a simple estimation for 
the effect of galaxy clustering that
was neglected in the Monte Carlo simulations. It should be pointed out that,
different from \citet{Patton2000} and \citet{Bell2006},
we assumed that the boundary separating
physical (i.e. gravitationally bound) and unphysical pairs, $\rm r_p$, 
is much larger than $\rm r_1 = 20\; h^{-1}\; kpc$, the separation limit 
in our pair selection. This is because the merger
time scales derived by \citet{Lotz2010}, as adopted in this work (Section 5.2),
are for close pairs with {\bf projected} separation $\rm r \leq 
20\; h^{-1}\; kpc$. 

\begin{figure}[!htb]
\plotone{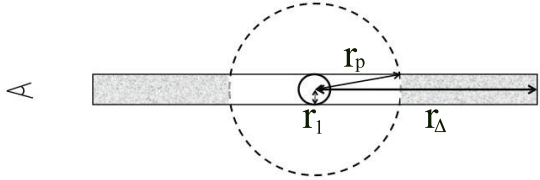}
\caption{Illustraction for Eq.~\ref{eq:eta1}. $\rm r_1$ is the maximum 
projected separation in the pair selection, $\rm r_p$ the outer boundary
of physical (i.e. gravitationally bound) pairs, 
and $\rm {r_\Delta}$ the distance range along the line-of-sight that
corresponds to the pair selection criterion for the photo-z difference:
$\rm |\Delta z_{phot}|/(1+z_{phot}) \leq 0.03$. The shadowed areas are
places where spurious companions are located.
}
\label{fig:projection}
\end{figure}

For a given two-point correlation function
$\rm \xi (r) = (r_0/r)^\gamma$, the additional SPF 
($\rm \eta$) due to clustering can be estimated as follows:
\begin{equation}
\eta = {4\pi n \int_{r_p}^{r_\Delta} (r_0/r)^\gamma r^2 dr 
\int_0^{arcsin(r_1/r)} sin(\theta)d\theta \over
 4\pi n \left[\int_0^{r_1} (r_0/r)^\gamma r^2 dr + \int_{r_1}^{r_\Delta} (r_0/r)^\gamma r^2 dr 
\int_0^{arcsin(r_1/r)} sin(\theta) d\theta \right] }. \label{eq:eta1}
\end{equation}
The numerator on the right-hand side of the equation is the probability to find 
spurious companions near a galaxy in both foreground and background
(in the shadowed areas in Fig.~\ref{fig:projection}).
Here $\rm {r_\Delta}$ is $\rm \sim 100\; Mpc$,
corresponding to the pair selection criterion of 
$\rm |\Delta z_{phot}|/(1+z_{phot}) \leq 0.03$.
The denominator is the probability of finding both real and spurious companions
with a projected separation of $\rm r \leq r_1$. 
The relations between parameters $\rm r_1$, $\rm r_p$ and $\rm r_\Delta$ 
are illustrated in  Fig.~\ref{fig:projection}. For $\rm \gamma = 2$ and
$\rm r_1 << r_p<< r_\Delta$, Eq.~\ref{eq:eta1} can be approximated by: 
\begin{equation}
\eta =  {r_1/r_p \over \pi}. \label{eq:eta2}
\end{equation}
Assuming $\rm r_p = 100\; h^{-1}\; kpc$, Eq.~\ref{eq:eta2} gives
$\rm \eta = 0.06$. It is worth noting that: (1) $\rm \eta$
is comparable to the SPF found by the Monte Carlo
simulations for the random associations; (2) 
$\rm \eta$ is constant against the redshift; 
(3) given the uncertainties in $\rm r_p$ and $\rm \gamma$, we shall
assume an relatively large error of 0.05 for $\rm \eta$;
(4) this correction also applies to pairs selected spectroscopically because
the condition $\rm {r_\Delta} >> r_p $ is still valid even for 
the selection criterion of
$\rm \Delta v \leq 500\; km\; sec^{-1}$.

\subsection{Fraction of Physical Pairs with $\rm \Delta v >
500\; km\; sec^{-1}$} 
We excluded physical pairs with $\rm 500
< \Delta v \lsim 1000\; km\; sec^{-1}$ from the merger rate
analysis since they
have very uncertain and long merger time scales. 
These high $\rm \Delta v$ pairs,
found in the group/cluster environments \citep{Domingue2009},
are not included in the above estimates for
projected unphysical pairs.
Here we make a separate correction for them.
Using local pairs taken from
\citet{Domingue2009}, selected using nearly identical
selection criteria as the CPAIR sample (cf. Section 4) except for that
spec-z were used and they include all pairs (isolated or in
groups/clusters) with $\rm \Delta v_{spec} < 1000$ km sec$^{-1}$, we
find that the fraction of physical pairs with $\rm \Delta v > 500\; km\;
sec^{-1}$ is $\rm 9.6\pm 3.0\%$ (Appendix C).

\subsection{Completeness and Reliability of the CPAIR Sample}
In Table~\ref{tbl:correction} 
we listed estimates of the correction factors for subsamples in
the four redshift bins. The combined completeness 
correction factor is the product of that due to missing very close pairs
and that due to photo-z errors.
The combined reliability correction factor, defined as $\rm 1 - SPF$,
is the product of that due to random projections,
the additional correction due to the clustering effect,
and that due to the contamination of physical pairs with 
$\rm \Delta v > 500\; km\; sec^{-1}$.
The combined completeness 
correction factor varies in the range of 0.60 -- 0.78 between the 4
redshift bins. The combined reliability
correction factor ($\sim 0.79$) is rather constant against the redshift.

As an independent check, exploiting a
sample of spec-z pairs (the ZPAIR sample) selected from the zCOSMOS
survey \citep{Lilly2007}, we made an empirical analysis on the
completeness and reliability of the CPAIR sample.
This resulted in an estimate of $\rm 0.86\pm
0.12$ for the completeness correction factor due to photo-z errors
(the incompleteness due to missing very close pairs cannot be checked
with spec-z pairs), which is consistent (within 1-$\sigma$) with
the result of the Monte Carlo simulations (Table~\ref{tbl:correction}).
The estimate for the reliability correction is
also $\rm 0.86\pm 0.12$, again consistent (within 1-$\sigma$) with
the values of the combined reliability correction in
Table~\ref{tbl:correction}.
The details of the analysis are presented in Appendix~D.

\setcounter{table}{3}
\tabletypesize{\scriptsize}
\begin{deluxetable*}{ccccccccccccccc}
\setlength{\tabcolsep}{0.05in} 
\tablenum{4}
\tablewidth{0pt}
\tablecaption{Mass Dependent Pair Fractions \label{tbl:fpair}}
\tablehead{
 \colhead{} 
  &  \multicolumn{2}{c}{$\rm z=0$} 
& &  \multicolumn{2}{c}{$\rm 0.2\leq z \leq 0.4$} 
& &  \multicolumn{2}{c}{$\rm 0.4< z \leq 0.6$} 
& &  \multicolumn{2}{c}{$\rm 0.6< z \leq 0.8$} 
& &  \multicolumn{2}{c}{$\rm 0.8 < z \leq 1$} 
\\
\cline{2-3} 
\cline{5-6} 
\cline{8-9} 
\cline{11-12} 
\cline{14-15} 
\\
\colhead{Mass Bin} 
  & \colhead{$\rm f_{pair}$} & \colhead{$\rm N_{pg}/N_{G}$}
& & \colhead{$\rm f_{pair}$} & \colhead{$\rm N_{pg}/N_{G}$}
& & \colhead{$\rm f_{pair}$} & \colhead{$\rm N_{pg}/N_{G}$}
& & \colhead{$\rm f_{pair}$} & \colhead{$\rm N_{pg}/N_{G}$}
& & \colhead{$\rm f_{pair}$} & \colhead{$\rm N_{pg}/N_{G}$}
}
\startdata
$9.4 \leq log(M)\leq  9.8$ & $0.014\pm  0.011$&   2/ 126 & &  $0.027\pm  0.006$ &  44/2034& &   ......        &  0/   0 & &  ......       &   0/   0 & &  ......        &   0/   0\\
$9.8 < log(M)\leq 10.2$    & $0.011\pm  0.006$&   7/ 524 & &  $0.020\pm  0.006$ &  24/1482& &   $0.030\pm  0.006$ & 44/1977 & &  ......       &   0/   0 & &  ......        &   0/   0\\
$10.2 < log(M) \leq 10.6$  & $0.013\pm  0.002$&  41/2775 & &  $0.043\pm  0.010$ &  39/1174& &   $0.030\pm  0.007$ & 34/1524 & &  $0.038\pm  0.007$&  71/2542 & &  ......        &   0/   0\\
$10.6 < log(M) \leq 11.0$  & $0.014\pm  0.002$&  97/5826 & &  $0.035\pm  0.011$ &  19/ 706& &   $0.025\pm  0.007$ & 18/ 955 & &  $0.044\pm  0.008$&  62/1913 & &  $0.045\pm  0.010$ & 110/3229\\
$11.0 < log(M) \leq 11.4$  & $0.011\pm  0.002$&  68/4520 & &  $0.042\pm  0.020$ &   6/ 183& &     (0)             &  0/ 267 & &  $0.021\pm  0.008$&  10/ 645 & &  $0.061\pm  0.015$ &  39/1051\\
$11.4 < log(M) \leq 11.8$  & $0.014\pm  0.007$&   8/ 417 & &    (0)             &   0/  17& &     (0)             &  0/  29 & &    (0)            &   0/  65 & &  $0.048\pm  0.029$ &   3/ 103\\
    total                  & $0.013\pm  0.001$& 223/14188 & & $0.030\pm  0.010$ & 132/5596& &   $0.026\pm  0.007$ & 96/4752 & &  $0.038\pm  0.009$& 143/5165 & &  $0.056\pm  0.011$ & 152/4383\\
\hline
\enddata
\end{deluxetable*}

\section{Local Pair Sample}
Pair statistics in the local universe were carried out 
using an updated version of the KPAIR sample by \citet{Domingue2009}, 
a close major-merger pair sample selected in the K-band from
cross matches between 2MASS and SDSS-DR5 galaxies.  
The update includes following modifications:
\begin{description}
\item{(1)} Stellar masses
of galaxies in KPAIR and in its parent sample are multiplied by a factor of 
$10^{-0.39}$. This is the average difference
between the mass estimated using the total $\rm K_s$ band
luminosity and a Salpeter IMF \citep{Domingue2009}, 
and the mass estimated using a Kroupa IMF by 
\citet{Kauffmann2003}. 
Because the mass estimated using the Kroupa IMF and that using
the Chabrier IMF are nearly identical \citep{Kauffmann2003}, this modification
makes the masses in the local KPAIR sample and those
in the CPAIR sample consistent. 
\item{(2)} In order to avoid possible bias due to
the local over-density (associated with the local super-cluster), a lower
redshift cut-off of $\rm v \geq 2000$ km sec$^{-1}$ ($\rm z \geq 0.0067$)
is introduced.
\item{(3)} Pairs with $\rm 500 < \Delta v \leq 1000\; km\; sec^{-1}$ 
are excluded.
Most of these pairs are in cluster environment, and may not be gravitationally 
bound. Excluding them improves the accuracy of the merger rate estimate.
\item{(4)} The magnitude limit is set at $\rm K_{s} = 12.5$, 
the completeness limit of the KPAIR sample.
\end{description}

There are 18,081 galaxies in the parent sample that are brighter than
$\rm K_{s} = 12.5$, of which 14,813 have measured redshifts (redshift
completeness of $\rm B_{z-comp}=0.82$), and 14,218 are in the range of
$0.0067\leq z \leq 0.1$.  The new paired galaxies sample has 221
galaxies, all brighter than $\rm K_{s} = 12.5$. Among them 188 are in
pairs with two measured redshifts of 
$\rm \Delta v_{spec} \leq 500\; km\; sec^{-1}$. 
These redshifts are in the range of $0.0067\leq z \leq
0.1$, with the median of z=0.042. The remaining 33 galaxies are in
single-redshift pairs (i.e. only one of the component galaxies having
measured redshift).

\section{Mass Dependent Merger Rates}
\subsection{Pair fraction}
The local pair fraction is calculated using the following formula:
\begin{equation}
\rm f_{pair,0} = {B_{z-comp}  \times (1 - \eta) \over A_0}\times {N_{2z} + N_{1z}\times 
    (1-Q_{spurious})\over N_{G,0}},
\end{equation}
where $\rm B_{z-comp}=0.82$ is the redshift completeness of the parent sample,
$\rm A_0 = 0.89$ the completeness of the local pair sample
\citep{Domingue2009}, $(1 - \eta) = 0.94$ the clustering related 
reliability correction factor found in Section 3.3.
$\rm N_{2z}$ and $\rm N_{1z}$ are numbers of galaxies in pairs of two 
measured redshifts and single redshifts, respectively,
$\rm N_{G,0}$ the number of galaxies in the parent sample
with measured redshift (in the range of $\rm 0.0067 \leq z <0.1$).
$\rm Q_{spurious} = 0.2$ is the probability for a single-redshift
pair to be a spurious pair \citep{Domingue2009}.
Note that $\rm A_0 \neq B_{z-comp}^2$ because pairs with
single measured redshifts were included. Also, we supplemented the 
SDSS redshifts of paired galaxies 
with redshifts found in the literature and in
our own redshift observations \citep{Domingue2009}.

The pair fractions in the COSMOS field are estimated as follows:
\begin{equation}
\rm f_{pair} = {Q_{reli}\over C_{comp}\times (1-D_{ACS})} 
\times {N_{pg}\over N_{G}},
\end{equation}
where $\rm Q_{reli}=[0.80, 0.79, 0.77, 0.78]$ is the reliability 
and $\rm C_{comp} = [0.78, 0.74, 0.71, 0.60]$
the completeness (for the four photo-z bins) of the pair sample, respectively;
$\rm D_{ACS} = 0.197$ is the fraction of pair candidates without
ACS images (not included in the final 
pair sample, see Section 2); $\rm N_{pg}$ is
the number of interacting galaxies in the pair sample, and $\rm
N_{G}$ the number of galaxies in the parent sample.

The relative error of the pair fraction (i.e. error/$\rm f_{pair}$)
can be estimated as the quadratic sum of
the random error $\rm \sigma_{rms}$ and the cosmic 
variance $\rm \sigma_{vari}$\footnote{The pair fraction, $\rm f_{pair}$,
is proportional to the probability of finding a
second galaxy within a spatial separation $r$ from a given galaxy: 
$\rm P(r) = 4\pi n \int^{r}_0 \, [1+\xi (r)] r^2 dr$, 
where $n$ is the number density of galaxies 
and $\rm \xi$ the two-point correlation function. 
Hence $\rm f_{pair}$ is proportional to $n$, and therefore
the cosmic variance has the same effect on $\rm f_{pair}$ as on the
number density.}:
\begin{equation}
\rm \sigma^2 = \sigma_{rms}^2 + \sigma_{vari}^2.
\end{equation}
The random (binomial statistics) error is
\begin{equation}
\rm \sigma_{rms}^2 =  {1-f_{pair}\over N_{pg}}.
\end{equation}
For the local sample, we adopted the approximation
$\rm N_{pg} = N_{2z} + N_{1z}(1-Q_{spurious})$. 
The cosmic variance is given by \citep{Peebles1980, Somerville2004}:
\begin{equation}
\rm \sigma_{vari}^2 = J_2(\gamma) \times (r_0/r_{samp})^\gamma  ,
\end{equation}
where $\rm r_0$ and $\gamma$ are the parameters in the two-point 
correlation function $\rm \xi (r) = (r_0/r)^\gamma$, 
$\rm r_{samp}$ the radius of the sampling volume,
and $\rm J_2$ a function of $\gamma$:
\begin{equation}
\rm J_2 = {72\over (3-\gamma)(4-\gamma)(6-\gamma)2^\gamma}.
\end{equation}
The correlation function parameters for local galaxies of different masses
were taken from \citet{Zehavi2005}. For galaxies of $z\geq 0.2$,
we assumed $\gamma = 1.8$ and derived the $\rm r_0$ values by
interpolating the measurements for galaxies
of different masses and redshifts by \citet{Zehavi2005},
\citet{Meneux2008} and \citet{Foucaud2010}. 
Cosmic variances for the integral pair fractions (i.e. not divided into
mass bins) were taken from that for galaxies in the mass bin of
$\rm 10.6 < \log(M_{star}/M_{\sun}) \leq 11$. 

\begin{figure}[!htb]
\plotone{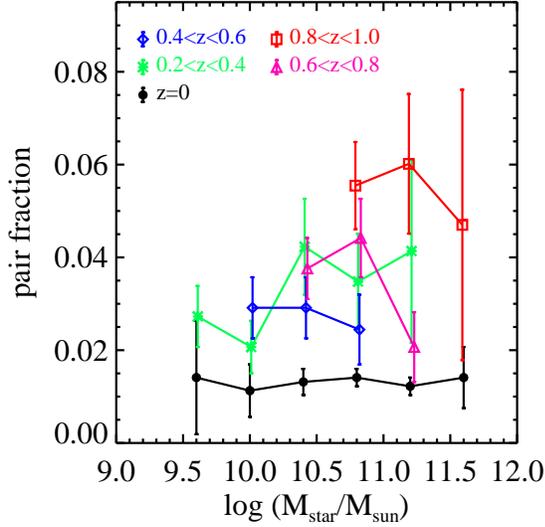}
\caption{Mass-dependent pair fractions in photo-z bins.}
\label{fig:fpair}
\end{figure}

The pair fractions so calculated are listed in Table~\ref{tbl:fpair} 
and plotted 
in Fig.~\ref{fig:fpair}.    
For local galaxies, we confirm the conclusion of \citet{Domingue2009}
that 
there is no significant mass dependence of the pair fraction. The integral 
pair fraction at z=0 is $\rm 1.3 \pm 0.1 \%$, slightly lower
than the result of \citet{Domingue2009} which is $\rm 1.6 \pm 0.1 \%$.
The difference is due to two factors: (1) the sample of 
\citet{Domingue2009} includes all 
pairs of $\rm \Delta v \leq 1000$ km sec$^{-1}$
whereas local pairs in this work are 
restricted to pairs of $\rm \Delta v \leq 500\; km\; sec^{-1}$;
(2) the application of the clustering related 
reliability correction factor ($(1-\eta) = 0.94$).

\begin{figure}[!htb]
\plotone{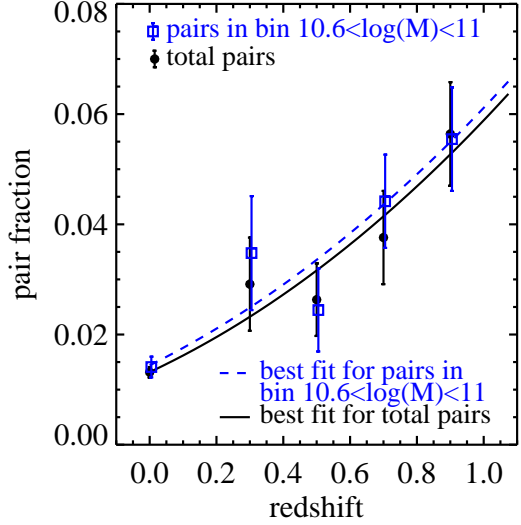}
\caption{Pair fraction evolution. The solid line is the least-square 
fit to total pair fraction (pair fraction of all galaxies regardless of
the stellar mass) vs. redshift relation, specified by
$\rm f_{pair} = 10^{-1.88(\pm 0.03)} (1+z)^{2.2(\pm 0.2)}$.  
The dashed line is the least-square 
fit to the pair fractions in the mass bin of
$\rm 10.6 < \log(M_{star}/M_{\sun}) \leq 11$, specified by
$\rm f_{pair} = 10^{-1.84(\pm 0.05)} (1+z)^{2.1(\pm 0.3)}$.  
}
\label{fig:fpairev}
\end{figure}

\begin{figure}[!htb]
\plotone{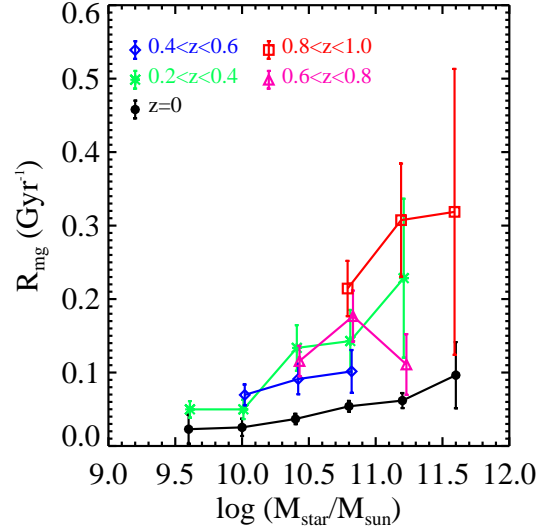}
\caption{Mass-dependent differential merger rate in 
different photo-z bins.
}
\label{fig:rmg}
\end{figure}

\begin{figure}[!htb]
\plotone{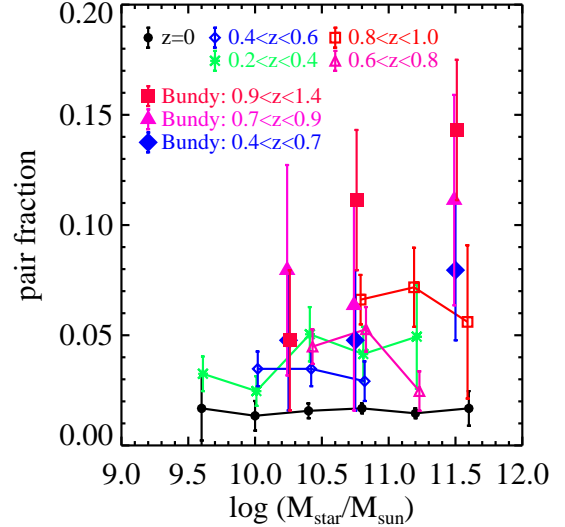}
\caption{Comparisons of mass-dependent pair fractions of this work
with those of \citet{Bundy2009}. Both results are converted to the 
$\rm f_{pair}$ for mergers of mass ratio, $\rm \mu$, $\leq 3$ (see text).}
\label{fig:mratbundy}
\end{figure}

There is no evidence for a significant mass dependence of the pair fractions
in higher photo-z bins, either. The trend for pair fractions to increase
with redshift can be seen in all mass bins, though with substantial scatter.
The major reason for the large scatter is the cosmic variance, given the
relatively small volume explored by the COSMOS survey in each photo-z bin
(Table~~\ref{tbl:parent}). 
In particular, there is a strong density enhancement in
the photo-z bin of $\rm 0.2 \leq  z < 0.4$ (D09), which biases
the pair fraction towards a higher value ($\rm f_{pair}$ is 
proportional to the density). Cosmic
variance often dominates the total error in the pair fraction: its 
contribution is usually $> 50\%$ except in
those bins where the number of paired galaxies, $\rm N_{pg}$, is less than 10
(hence the random error is large).
In Fig.~\ref{fig:fpairev}, we plot the redshift dependence
of the integral pair fraction (pair fraction of all galaxies regardless
of the stellar mass), and that of the pair fraction for galaxies 
in the mass bin $\rm 10.6 < \log(M_{star}/M_{\sun}) \leq 11$ (the bin
encompassing the $\rm M^*_{star}$). These two results are very close 
to each other,
in agreement with our conclusion that the pair fraction does not vary
significantly with stellar mass.

The least-square fit to the redshift
dependence of the integral pair fractions
is $\rm f_{pair} = 10^{-1.88(\pm 0.03)} (1+z)^{2.2(\pm 0.2)}$.  
For the pair fractions in 
the mass bin of $\rm 10.6 < \log(M_{star}/M_{\sun}) \leq 11$, the best fit is
$\rm f_{pair} = 10^{-1.84(\pm 0.05)} (1+z)^{2.1(\pm 0.3)}$.  

\subsection{Differential Major-Merger Rate}

\setcounter{table}{4}
\begin{deluxetable*}{ccccccccccc}
\tabletypesize{\footnotesize}
\setlength{\tabcolsep}{0.05in} 
\tablenum{5}
\tablewidth{0pt}
\tablecaption{Mass Dependent Differential Merger Rate \label{tbl:rmg}}
\tablehead{
 \colhead{Mass Bin} 
& & \multicolumn{9}{c}{$\rm R_{mg}$ (Gyr$^{-1}$)} \\
\cline{3-11}
\\
\colhead{} 
& &  \colhead{$\rm z=0$} 
& &  \colhead{$\rm 0.2\leq z \leq 0.4$} 
& &  \colhead{$\rm 0.4< z \leq 0.6$} 
& &  \colhead{$\rm 0.6< z \leq 0.8$} 
& &  \colhead{$\rm 0.8 < z \leq 1$} 
}
\startdata
$9.4 \leq log(M)\leq  9.8$&  & $0.023\pm 0.019$&&  $0.050\pm0.011 $ & &   ......         & &  ......        & &  ......        \\
$9.8 < log(M)\leq 10.2$   &  & $0.025\pm 0.012$&&  $0.049\pm0.013 $ & &   $0.068\pm 0.014$ & &  ......        & &  ......        \\
$10.2 < log(M) \leq 10.6$ &  & $0.037\pm 0.008$&&  $0.132\pm0.030 $ & &   $0.089\pm 0.020$ & &  $0.112\pm 0.020$& &  ......        \\
$10.6 < log(M) \leq 11.0$ &  & $0.055\pm 0.007$&&  $0.140\pm0.042 $ & &   $0.099\pm 0.029$ & &  $0.171\pm 0.033$& &  $0.211\pm 0.037$ \\
$11.0 < log(M) \leq 11.4$ &  & $0.062\pm 0.011$&&  $0.226\pm0.107 $ & &   ......           & &  $0.108\pm 0.040$& &  $0.302\pm 0.076$ \\
$11.4 < log(M) \leq 11.8$ &  & $0.096\pm 0.046$&&  ......           & & ......             & &  ......          & &  $0.313\pm 0.191$ \\
\hline
\enddata                                                                        
\end{deluxetable*}

The differential major-merger rate is the probability for each
galaxy to be involved in a major merger per Gyr: $\rm R_{mg} \propto
f_{pair}/T_{mg}$, where $\rm T_{mg}$ is the merger time scale in
Gyr. Because the physical process of a galaxy merger is
very complex (see Hopkins 2010b for a review),
$\rm T_{mg}$ has been a major source of uncertainty in the merger
rate studies. In the literature the most common approach has been the
approximation of $\rm T_{mg}$ by the dynamical friction time scale
\citep{Binney1987, Patton2000, Jiang2008, Kitzbichler2008}.
\citet{Kitzbichler2008} studied
the $\rm T_{mg}$ in a semi-analytical model built on the 
results of the Millennium Simulation. They
assumed circular orbits to estimate the dynamical friction process, 
and found relatively weak mass and redshift 
dependence in the form of $\rm T_{mg} \propto M_{star}^{-0.3}(1+z/8)$.
However, their dynamical friction time is only appropriate for
small satellite galaxies at large radii. For the massive close major-merger
pairs in our samples, it becomes a very poor approximation because of 
two issues \citep{Hopkins2010b}: (1) Angular momentum loss at these
radii is not dominated by the dynamical friction, but rather by exchange
in strong resonances between the baryonic components that act much more 
efficiently. (2) By these radii, even the initially circular orbits have
become highly radial, leading to shorter merger times.

On these reasons, we
instead estimated $\rm T_{mg}$ using the results of \citet{Lotz2010}. They
carried out
high resolution hydro-dynamical simulations for a large number 
of galaxy mergers
with diverse initial conditions, and derived $\rm T_{mg}$ at different
projected separations in a view-angle averaged format. These simulated mergers
have line-of-sight velocity difference $\rm \Delta v \leq 500\; km\; sec^{-1}$,
identical to the pairs in our sample. Nine mergers in \citet{Lotz2010} have 
baryonic mass ratios $\leq 3$ (i.e. one 1:1 merger and eight 3:1 mergers)
with $\rm M_{star} \geq 10^{10.7}\; M_\sun$.
In the bin of $\rm 5 \leq r_{proj} \leq 20\; h^{-1}\; kpc$, these nine mergers
have an average merging time scale of $\rm T_{mg} = 0.30\pm 0.06 \; Gyr$.
The three 1:1 mergers in \citet{Lotz2010} have stellar masses in the range
of $\rm 10^{9.7}$ -- $\rm 10^{10.7}\;  M_\sun$, and their $\rm T_{mg}$'s in the
$\rm 5 \leq r_{proj} \leq 20\; h^{-1}\; kpc$ bin show a weak mass dependence 
of  $\rm T_{mg} \sim M_{star}^{-0.2}$, consistent with what found by
\citet{Kitzbichler2008}. Because \citet{Lotz2010} did not study the 
red-shift dependence of $\rm T_{mg}$, and because their sample is too small
(3 mergers) to derive a meaningful mass dependence, we adopt the
relation found by \citet{Kitzbichler2008}, namely 
$\rm T_{mg} \propto M_{star}^{-0.3}(1+z/8)$. 
These dependences are sufficiently weak that the associated 
uncertainties will not have any significant effect on our results.
The final merger time scale we adopted is:
\begin{equation}
\rm T_{mg} = 0.3 \; Gyr \times \left(
    {M_{star} \over 10^{10.7} M_{\sun}} \right)^{-0.3}
    \left(1+{z\over 8} \right)     \label{eq:Tmg}.
\end{equation}
And the differential merger rate for major mergers of mass ratio $\leq 3$
is:
\begin{equation}
\rm R_{mg} = A \times f_{pair} / T_{mg},
\end{equation}
where $\rm A = 1.19$ is the factor converting the pair fraction
in this work (for mergers of mass ratio $\rm \mu \leq 2.5$) to that
of mergers of mass ratio $\rm \mu \leq 3$. Here we assumed that
$\rm f_{pair} \propto \log(\mu_{max})$ (Appendix~E).
In Table~\ref{tbl:rmg} and Fig.~\ref{fig:rmg} 
we present our results on $\rm R_{mg}$.

Using the least-square fit to the pair fraction evolution, 
$\rm f_{pair} = 10^{-1.88(\pm 0.03)} (1+z)^{2.2(\pm 0.2)}$,
and Eq.~\ref{eq:Tmg}, we derived a best-fit function for
the mass-dependent $\rm R_{mg}$ evolution:
\begin{equation}
\rm R_{mg} (M_{star},z) = 0.053\times
\left( {M_{star} \over 10^{10.7} M_{\sun}} \right)^{0.3}
                      {(1+z)^{2.2}\over 1+z/8} 
                                  \; (Gyr^{-1})    \label{eq:Rmg}.
\end{equation}
Integrating this merger rate over time, we find that the
probability for individual
galaxies to be involved in a major merger since z=1 is 
0.8$\rm \times (M_{star}/10^{10.7} M_{\sun})^{0.3}$. Accordingly, on average,
massive galaxies of $\rm M_{star} \sim 10^{10\hbox{--} 11.5} M_{\sun}$ have 
undergone $\sim 0.5$ -- 1.5 times mergers since z=1.

\subsection{Comparisons with Previous Results}
\begin{figure}[!htb]
\plotone{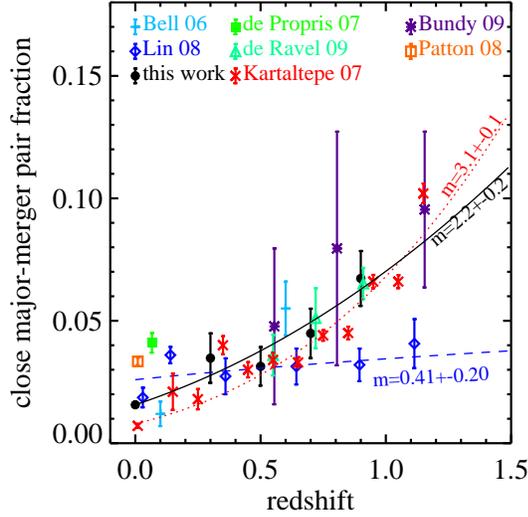}
\caption{Comparisons of observed pair fractions in the
  literature. When it is appropriate, results of different 
  authors were corrected so they 
  are consistent with a common definition of close
  major-merger pairs with the maximum projected
  separation of $\rm r_{proj,max} = 20\; h^{-1}\; kpc$ 
 and the maximum primary-to-secondary mass ratio $\rm \mu_{max} = 3$. 
  Results 
  of pair samples of different $\rm r_{proj,max}$ were corrected
  by assuming $\rm f_{pair} \propto (r_{proj,max})^{3-\gamma}$ with
  $\gamma=2$ \citep{Bell2006}. 
  For example, results of \citet{Lin2008} were divided
  by a factor of $\rm 1.5$ because they had $\rm
  r_{proj,max} = 30\; h^{-1}\; kpc$.  The results of \citet{Bell2006}
  had $\rm r_{proj,max} = 30\; kpc$, corresponding to $\rm
  r_{proj,max} = 21\; h^{-1}\; kpc$ for $h=0.7$, very close to the $\rm
  r_{proj,max} = 20\; h^{-1}\; kpc$ and therefore no correction was
  applied.  Results of pair samples of different mass ratio limits
  were corrected by assuming $\rm f_{pair} \propto \log(\mu_{max})$
  (Fig.~\ref{fig:mratdist}).  
  These include results of this work ($\rm \log(\mu_{max})
  =0.4$), \citet{Bundy2009} and \citet{deRavel2009} (both having $\rm
  \log(\mu_{max}) =0.6$), and \citet{Patton2008} ($\rm \log(\mu_{max})
  =0.3$).  This correction was not applied to results for pair
  samples without the mass ratio cutoff (e.g. K07, 
  \citealt{Bell2006}, and \citealt{dePropris2007}). 
}
\label{fig:mratplot}
\end{figure}

Our results on the mass independence of the local pair fraction 
(filled circles in Fig.~\ref{fig:fpair}) are
in agreement with those of \citet{Domingue2009} and \citet{Patton2008}
while contradicting \citet{Xu2004}, the latter were derived
using a small sample of 19 pairs. 
For pairs of higher redshifts, \citet{Bundy2009}
found a trend of positive mass dependence, 
which was not confirmed by our results.
In Fig.~\ref{fig:mratbundy}, our results 
are compared to those of \citet{Bundy2009}. 
In order to compensate the difference in the 
mass ratios in the two works ($\rm \mu \leq 2.5$ in this work and
$\rm \mu \leq 4.0$ in \citealt{Bundy2009}), 
both results are converted to the 
$\rm f_{pair}$ for $\rm \mu \leq 3$ mergers.
In Appendix~E, it is shown that the pair fraction 
$\rm f_{pair}$ increases proportionally with
$\log (\mu_{max})$. Accordingly,
the pair fractions from this work
were scaled up by a factor of $\rm 1.19 = \log(3)/0.4$, and those of
\citet{Bundy2009} were scaled down by a factor of 
$\rm 0.80 = \log(3)/0.6$.
The results of \citet{Bundy2009}
might have suffered from large uncertainties: 
Those obtained using their ``method I'' 
(projected pairs without any redshift information for the companions),
which are plotted in Fig.~\ref{fig:mratbundy}, were based on pair samples with
high contaminations ($\sim $60 -- 70\%)
of unphysical pairs; and those from their ``method II'' (spectroscopic
and/or photometric redshifts for both components) were based on small
samples (3 to 15 paired galaxies in each mass/redshift bin).
\citet{deRavel2009} claimed evidence for strong mass 
dependence of the evolutionary index of the pair fraction, in the sense
that low mass pairs have strong pair fraction evolution ($\rm m = 3.13\pm
1.54$) and high mass pairs have weak evolution ($\rm m = 0.52\pm
2.07$). However their results are very uncertain, as indicated by their
large errors. \citet{deRavel2009} also claimed evidence for strong
evolution ($\rm m = 4.73\pm 2.01$) in optically faint pairs ($\rm M_B \leq
-18 - Qz$, Q=1.11) and for weak evolution 
($\rm m = 1.50\pm 0.76$) in optically bright pairs ($\rm M_B \leq
-18.77 - Qz$). But the low evolutionary index of the bright pairs was obtained
only when they included in their fit 
the z=0 pair fraction of \citet{dePropris2007}, one of the highest
local pair fraction in the literature (Fig.~\ref{fig:mratplot}). 
Indeed, when being calculated in
the same way as for the evolutionary index of faint pairs (i.e. fitting only
high z data points), the index of 
bright pairs is $\rm m = 3.07\pm 1.68$, consistent with that for faint pairs.

\begin{figure}[!htb]
\plotone{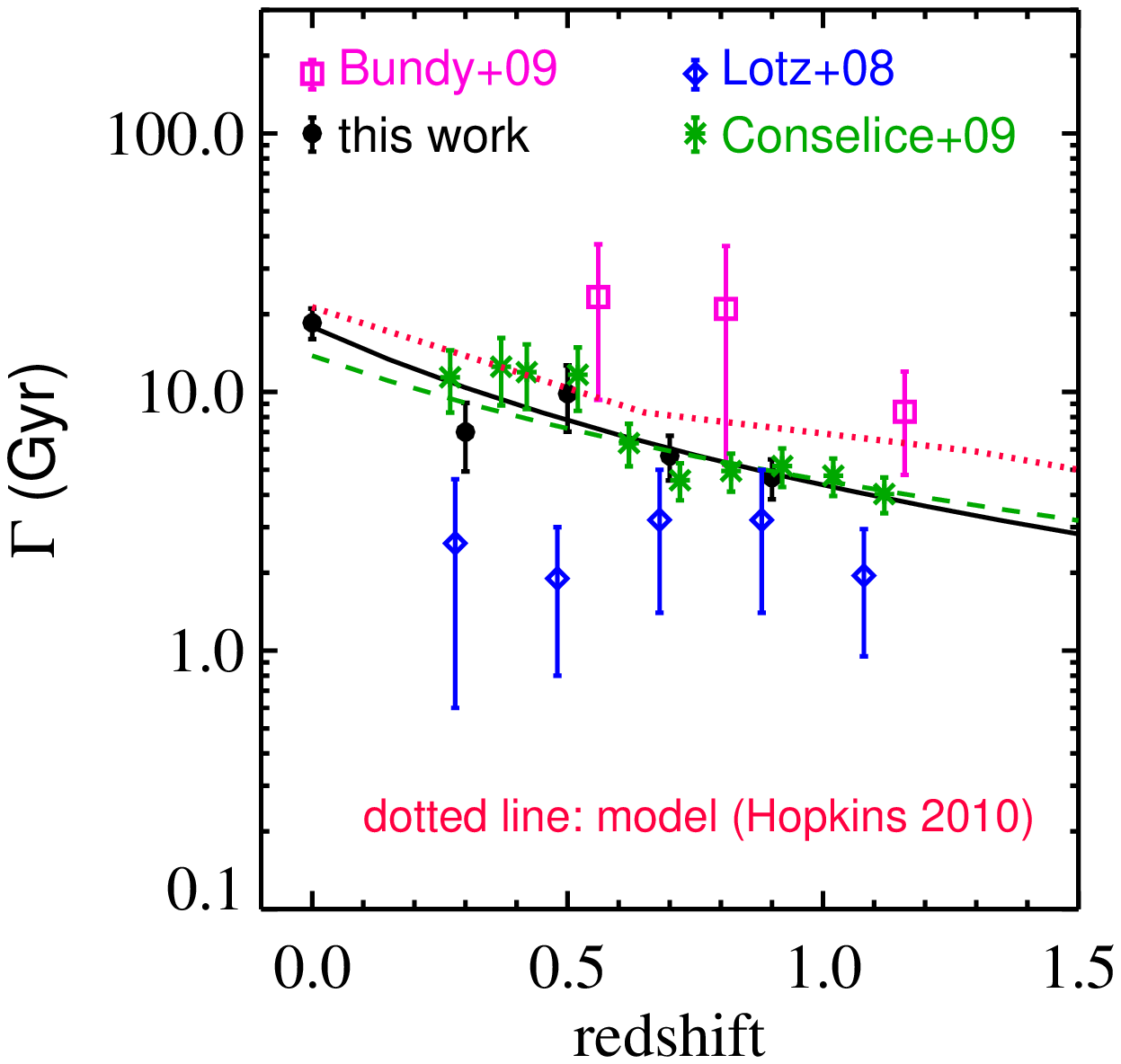}
\caption{The redshift dependence of $\rm \Gamma = 1/R_{mg}$.
The values of morphological based merger studies of 
\citet{Conselice2009} and of \citet{Lotz2008} were taken
from Fig.~7 of \citet{Conselice2009}. $\rm R_{mg}$'s of 
\citet{Bundy2009} were scaled down by a factor of $\rm 0.80 = \log(3)/0.6$
to make it consistent with the merger rate for $\mu \leq 3$ mergers.
The solid line is the inverse
of Eq.~\ref{eq:Rmg} of this work. The dashed line is the
best fit of \citet{Conselice2009}. The dotted line is the model prediction
of \citet{Hopkins2010a}.
}
\label{fig:gamma}
\end{figure}
\begin{figure}[!htb]
\plotone{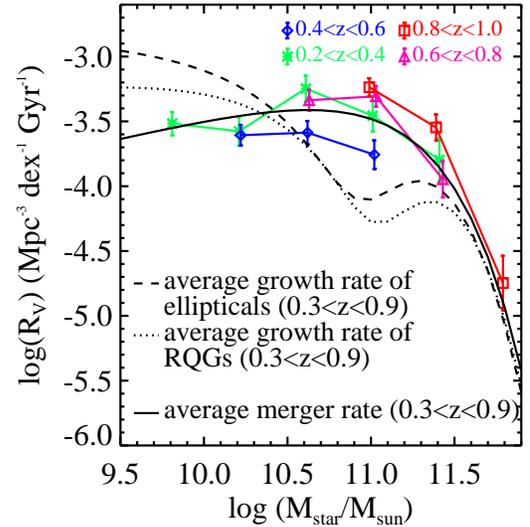}
\caption{Data points with error bars are mass-dependent volume merger rates 
in different redshift bins. The dashed line is the average growth rate 
of ellipticals between z=0.3 -- 0.9 and dotted line that
of RQGs, estimated using
the mass function of \citet{Ilbert2010}. The solid line 
is the average volume merger rate calculated using Eq.~\ref{eq:R_V} by 
replacing $\rm R_{mg}$ with its best fit (Eq.~\ref{eq:Rmg}) and
averaged over the same redshift range of 0.3 -- 0.9.
}
\label{fig:rv}
\end{figure}

In Fig.~\ref{fig:mratplot}, our results on the cosmic evolution of the
integral pair fraction 
are compared with those taken from the literature.
It shows that the evolutionary rate derived from our results
is in between those for the strong evolution (e.g.
the result of K07) and for
weak evolution (e.g. the result of \citealt{Lin2008}), respectively. 
Actually, our pair fractions in the photo-z bins of $\rm z=0.2$--1.0 agree
well with those of K07 in the same redshift range.
The marginally significant 
difference between the two evolutionary rates is mainly due to the 
relatively low pair fraction at z=0,
which may be caused by an incompleteness associated to the
``missing secondary'' bias \citep{Xu2004}, and
the relatively high pair fraction at z=1.3 in the results of 
K07. We clearly see much stronger evolution than that of 
\citet{Lin2008}. Their results,
based on spectroscopically confirmed pair samples in incomplete redshift
surveys, may have relatively large statistical uncertainties because of the
large correction factors (a factor of $>3$) for the incompleteness.
The color based pre-selection of their redshift surveys may indeed introduce
biases in the pair selection, given the significant influence of
galaxy-galaxy interaction on optical colors \citep{Larson1978}.

\setcounter{table}{5}
\begin{deluxetable*}{ccccccccccc}
\tabletypesize{\footnotesize}
\setlength{\tabcolsep}{0.05in} 
\tablenum{6}
\tablewidth{0pt}
\tablecaption{Mass Dependent Volume Merger Rate \label{tbl:rv}}
\tablehead{
 \colhead{Mass Bin} 
& & \multicolumn{9}{c}{$\rm \log(R_{V})\; (Mpc^{-3}dex^{-1}Gyr^{-1})$} \\
\cline{3-11}
\\
\colhead{} 
& &  \colhead{$\rm z=0$} 
& &  \colhead{$\rm 0.2\leq z \leq 0.4$} 
& &  \colhead{$\rm 0.4< z \leq 0.6$} 
& &  \colhead{$\rm 0.6< z \leq 0.8$} 
& &  \colhead{$\rm 0.8 < z \leq 1$} 
}
\startdata
$9.4 \leq log(M)\leq  9.8$&  & $-3.81\pm 0.28$&&  $-3.51\pm 0.09$ & &   ......         & &  ......        & &  ......        \\
$9.8 < log(M)\leq 10.2$   &  & $-3.86\pm 0.18$&&  $-3.57\pm 0.11$ & &  $-3.61\pm 0.08$ & &  ......        & &  ......        \\
$10.2 < log(M) \leq 10.6$ &  & $-3.81\pm 0.10$&&  $-3.25\pm 0.10$ & &  $-3.59\pm 0.09$ & & $-3.34\pm 0.08$& &  ......        \\
$10.6 < log(M) \leq 11.0$ &  & $-3.83\pm 0.09$&&  $-3.45\pm 0.12$ & &  $-3.77\pm 0.11$ & & $-3.31\pm 0.08$& & $-3.24\pm 0.07$ \\
$11.0 < log(M) \leq 11.4$ &  & $-4.13\pm 0.09$&&  $-3.79\pm 0.17$ & &   ......         & & $-3.95\pm 0.14$& & $-3.55\pm 0.10$ \\
$11.4 < log(M) \leq 11.8$ &  & $-4.74\pm 0.18$&&  ......          & &   ......         & & ......         & & $-4.75\pm 0.21$ \\
\hline
\enddata
\end{deluxetable*}

In the literature, pair fractions are often compared to 
merger fractions estimated using counts of peculiar galaxies
\citep{Conselice2003, Conselice2009, Lotz2008, Jogee2009}.
In general the latter are higher than the former,
because (1) Contaminations from irregular galaxies \citep{Jogee2009};
(2) morphologically selected merger samples 
based on the G -- $\rm M_{20}$ method \citep{Lotz2008, Lotz2010}
include minor mergers; 
(3) the merger time scales for morphologically selected merger samples 
based on the CAS method are
longer than the merger time scales  of close 
 major-merger pairs \citep{Conselice2009}.
Given the different merger time scales for close pairs and for
peculiar galaxies, it is more appropriate to compare the
differential merger rates $\rm R_{mg}$. In Fig.~\ref{fig:gamma}
 we compare 
the inverse of the
$\rm R_{mg}$, $\rm \Gamma = 1/R_{mg}$ \citep{Conselice2009},
of morphologically selected mergers by \citet{Conselice2009} and by
\citet{Lotz2008} with that of $\rm M^*_{star}$ galaxies 
($\rm \log(M^*_{star}/M_\sun) \sim 10.8$, \citet{Ilbert2010}) 
in close major-merger pairs in this work and in \citet{Bundy2009}.
The $\rm \Gamma$ parameter derived using paired galaxies
in our sample (the inverse of Eq.~\ref{eq:Rmg}) and that of
\citet{Conselice2009} derived using morphologically selected mergers
are in very good agreement. The higher $\rm \Gamma$ values of \citet{Bundy2009} 
are likely due to the relatively long merging time scale they
adopted from \citet{Kitzbichler2008}. On the other hand, the
low $\rm \Gamma$ values and lack of evolution
of \citet{Lotz2008} are because of the inclusion
of minor mergers in their sample. Our results
are in good agreement with the prediction of the default (semi-empirical)
model of \citet{Hopkins2010a}.

\section{Major Mergers, Elliptical Galaxy Formation, and Galaxy Assembly}
\subsection{Mass Dependent Volume Merger Rate}

\begin{figure*}[!htb]
\plottwo{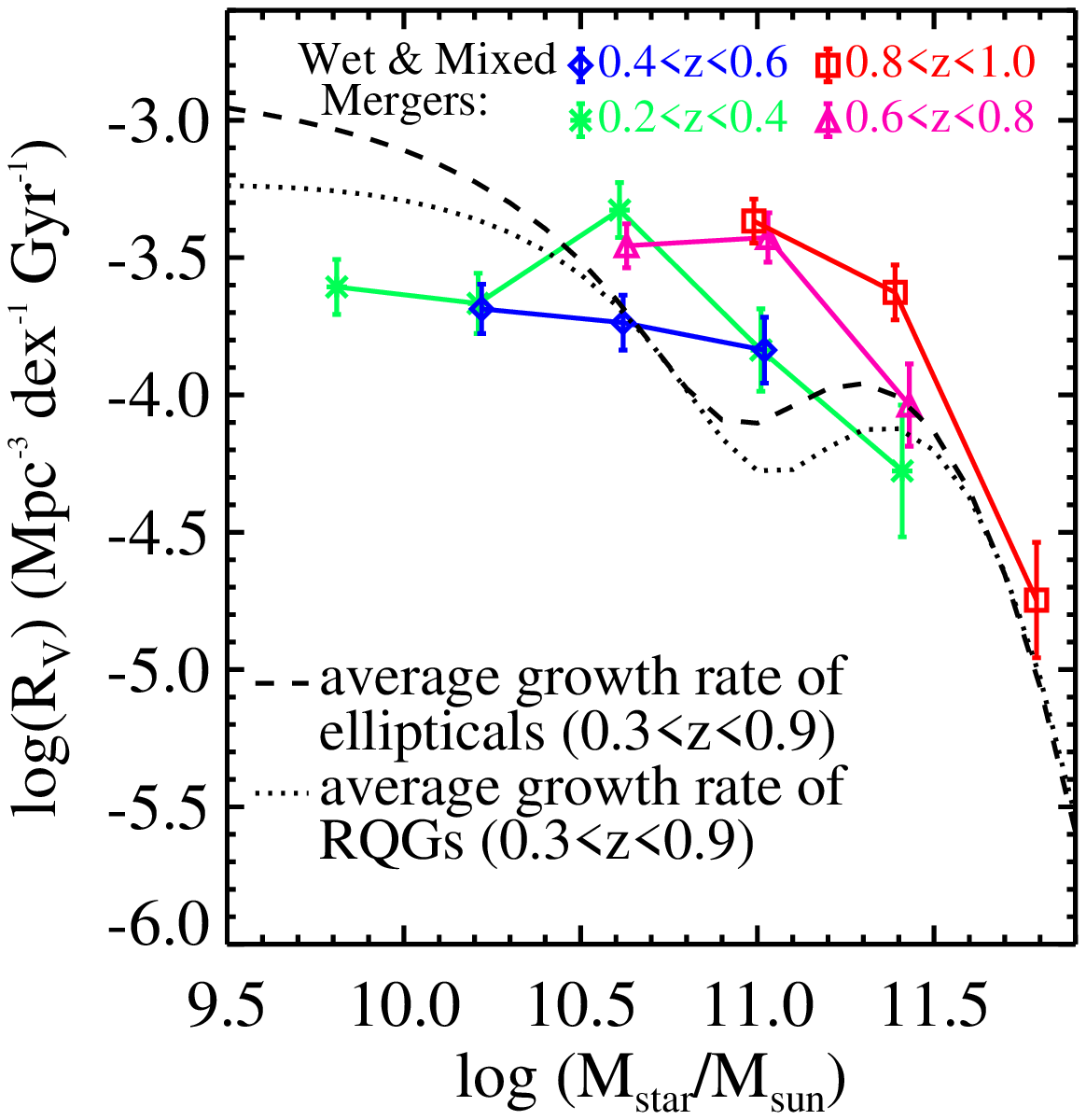}{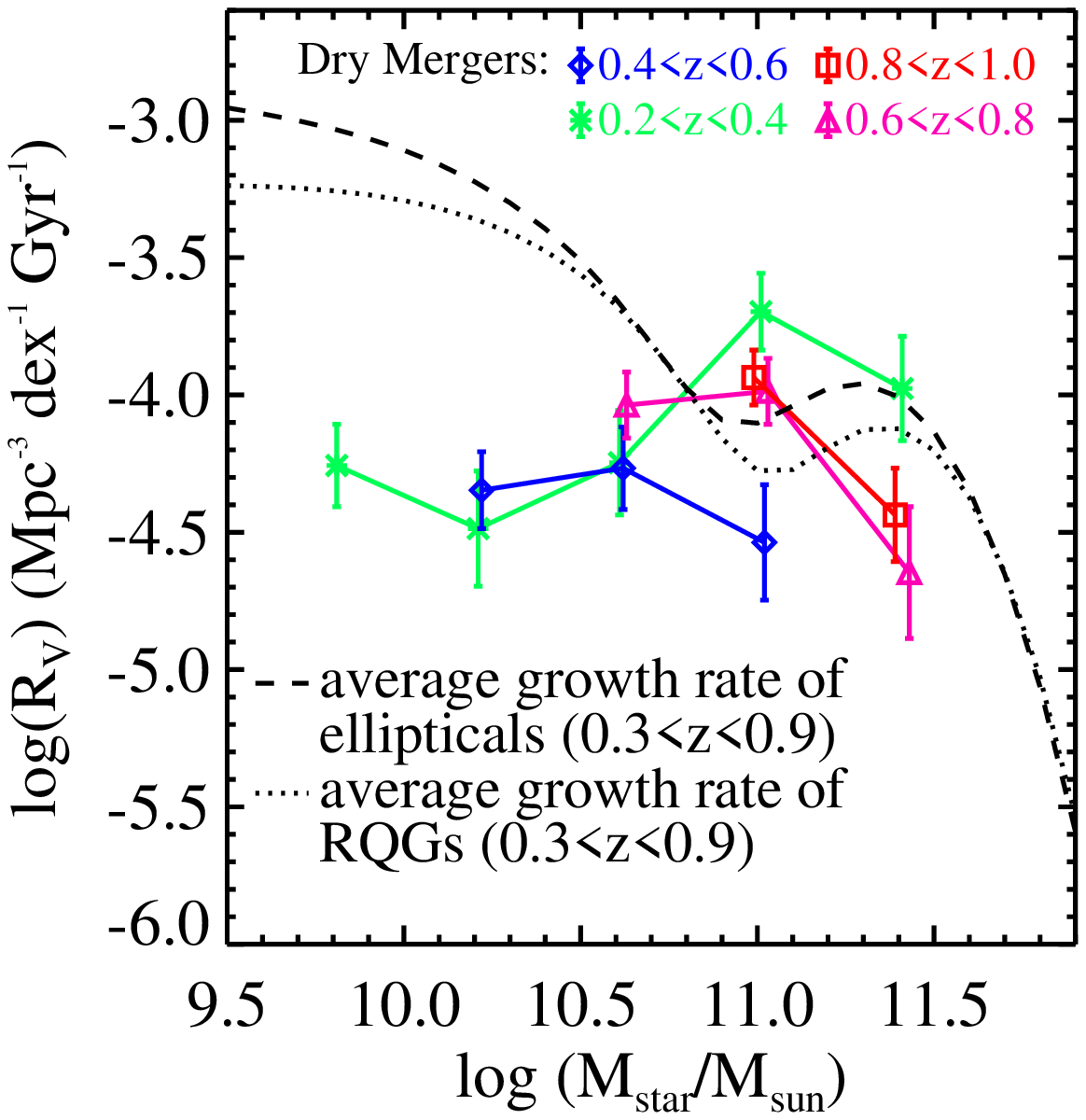}
\caption{
{\bf Left:} Mass-dependent volume merger rates of wet (S+S) and
mixed (S+E) mergers,
in different photo-z bins. The dashed line is the average growth rate 
of ellipticals between z=0.3 -- 0.9 and dotted line that
of RQGs, both being estimated using
results of \citet{Ilbert2010}. \break
{\bf Right:} Mass-dependent volume merger rates of dry (E+E) mergers.
}
\label{fig:wetdry}
\end{figure*}

The volume merger rate, $\rm R_{V}$, measures the frequency of merger
events in a given volume in the universe. Here we define the mass dependent
$\rm R_{V}$ in terms of the stellar mass of the {\it merger
remnant}, which is the total stellar mass of the two galaxies involved
in the merging (ignoring the mass of stars formed during the merger):
\begin{equation}
\rm R_{V} (M_{star},z) = 0.5 \times R_{mg} (M_{star}/10^{0.2},z)
\phi (M_{star}/10^{0.2},z)     \label{eq:R_V},
\end{equation}
where $\rm \phi (M,z) $ is the GSMF of galaxies in the parent sample
(D09), the factor of 0.5 is due to the fact that every major 
merger event involves two galaxies of similar mass. We also assume
that on average the mass of a merger remnant is 0.2 dex higher than
that of individual galaxies involved in the merger. This is because, under
the assumption that mass ratio distribution is flat (Appendix C), 
pairs in our sample has a mean mass ratio of 0.2 dex.
Therefore the logarithm of the mean ratio between 
the total mass of a pair and that of 
the primary is $\rm \log(1+10^{-0.2}) \simeq 0.2$. Our results on
the $\rm R_V$ are presented in Table~\ref{tbl:rv}.

In Fig.~\ref{fig:rv}
 we compare our results with
the average growth rate of elliptical galaxies (dashed line)
and that of RQGs (dotted line) between z=0.3 -- 0.9, 
estimated using the differences between their GSMFs at z=0.3 and z=0.9, 
taken from \citet{Ilbert2010}, divided by 3.88 Gyr
(the time span corresponding to the redshift interval of [0.3,0.9]).
The solid line 
is the average volume merger rate calculated using Eq.~\ref{eq:R_V} by 
replacing $\rm R_{mg}$ with its best fit (Eq.~\ref{eq:Rmg}) and averaged
over the same redshift range of 0.3 -- 0.9. It shows that major mergers
can fully account for the formation
of both massive ellipticals and RQGs 
($\rm M_{star} \geq 10^{10.5}\; M_\sun$). This contradicts
\citet{Bundy2009} who concluded that the major-merger rate is too low
to fully explain the formation of RQGs since z=1.
The major reason for the contradiction is due to the difference in the 
adopted merger time scales in this work and in \citet{Bundy2009}: 
Our $\rm T_{mg}$, derived from the results of \citet{Lotz2010}, is about
a factor of 2 shorter than that used by \citet{Bundy2009}. 
There is also a difference in the formation rates of ellipticals 
and RQGs adopted
in this work (estimated from results of \citet{Ilbert2010}) and
in \citet{Bundy2009}. The latter is about 50 -- 100\% higher than the former.

We define ``dry mergers'' (``wet mergers'') as those in pairs or
multiple systems consisted of only RQGs (SFGs), and ``mixed mergers''
the rest of galaxies in the pair sample. In Fig.~\ref{fig:wetdry}
we compare the volume
merger rates of dry mergers, and those of wet and mixed mergers
combined, to the formation of ellipticals. Our results show that
wet/mixed mergers alone can account for the formation rate of massive
ellipticals and RQGs, even for the most massive ones of $\rm M_{star} \geq
10^{11.3}\; M_\sun$.  Our results are consistent with \citet{Lin2008},
who also found that the wet and mixed mergers dominated over
the dry mergers since z$\sim 1$.

\begin{figure}[!htb]
\plotone{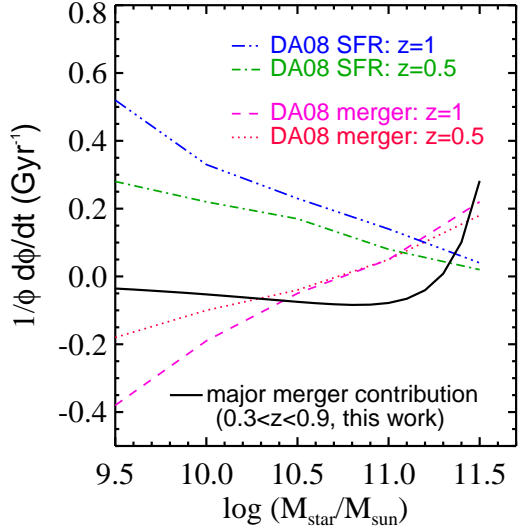}
\caption{Fractional change rate of the GSMF. The solid line is our
result on the mean $\rm \hat{\phi}_{merger}$ 
(contribution of major mergers, see Eq.~\ref{eq:phi}) over $\rm 0.3 < z < 0.9$.
The dotted and dashed lines are results of \citet{Drory2008} on
the contribution of mergers (including both major and minor mergers) at z=0.5
and z=1, respectively.
The dot-dashed and dot-dot-dot-dashed lines 
are results of \citet{Drory2008} on
the contribution of star formation at z=0.5
and z=1, respectively.}
\label{fig:gsmf}
\end{figure}
\begin{figure}[!htb]
\plotone{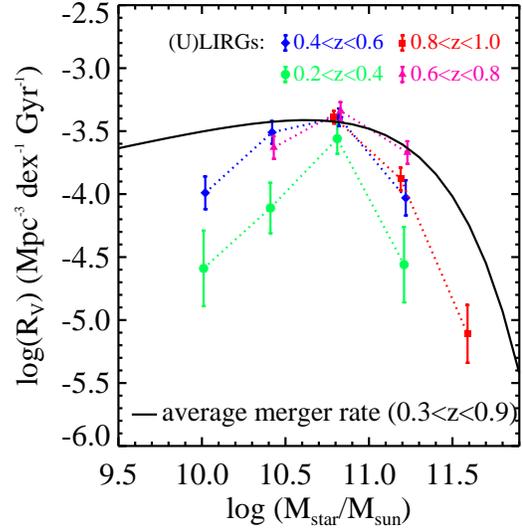}
\caption{The volume rates of (U)LIRGs
($\rm 10^{11.5} < L_{IR}/L_\sun < 10^{12.5}$), in different
redshift and stellar mass bins. Estimated using data taken from 
\citet{Kartaltepe2010} and assuming a (U)LIRG time scale of 140 Myr.
The solid line is the average mass dependent volume merger rate (this work),
identical to that in Fig.~\ref{fig:rv}. 
}
\label{fig:ulirg}
\end{figure}

It should be pointed out that 
ellipticals (Es) and RQGs are not identical (albeit with 
large overlap) since there are both red disk galaxies 
\citep{Bamford2009, Bundy2010} and blue ellipticals
\citep{Kannappan2009, Huertas-Company2010}. Both Es and RQGs
are likely originated from star forming disk galaxies: RQGs  
formed through SFR quenching \citep{Bell2007, Faber2007},
and ellipticals through mergers \citep{Toomre1978, Barnes1988} or
secular evolution \citep{Kormendy2004}. \citet{Hopkins2008}
argue that only major mergers can reproduce the kinematic properties
of massive ellipticals, whereas some 
low mass ellipticals are ``pseudo-bulges'' formed 
through disk instabilities and secular evolution in late type galaxies.

Many quenching mechanisms in
the literature are related either directly to merger induced
feed-backs (e.g. gas consumption by extreme starbursts and gas loss
due to superwinds) or to massive bulges (such as the AGN
quenching, \citealt{Benson2003, Somerville2008},
and morphological quenching, \citealt{Martig2009}). 
Hence a close relation
between major mergers and RQGs formation is expected. Indeed
\citet{Hopkins2008} argued that a wide range of observations
(e.g. the bivariate red fraction as a function of galaxy and halo
mass, the density of passive galaxies at high redshifts, the
emergence/evolution of the color-morphology-density relations at high
redshift, and the fraction of disky/boxy spheroids as a
function of mass) favor a major-merger related quenching model 
to other quenching models.

Our results (Fig.~\ref{fig:rv} and Fig.~\ref{fig:wetdry}) support
the model of \citet{Hopkins2008}.
Major mergers, dominantly wet or mixed, can fully account for
the formation rates of both ellipticals and RQGs with 
$\rm M_{star} \geq 10^{10.5}\; M_\sun$. For most
massive galaxies with $\rm M_{star} \geq 10^{11.3}\; M_\sun$, the major-merger
rate agrees very well with the two formation rates
(in this mass range most ellipticals and RQGs 
 belong to the same population of red elliptical galaxies).
In the mass range of $\rm 10^{10.5}\leq M_{star} \leq 10^{11.3}\; M_\sun$,
the major-merger rate is slightly higher
than both formation rates.
Two factors may be responsible for this: (1) Remnants of some gas-rich
wet mergers may remain to be blue disk galaxies \citep{Hopkins2009}.
(2) Dry mergers, contributing most in this mass 
range, may move some red ellipticals to higher mass.

Fig.~\ref{fig:rv} and Fig.~\ref{fig:wetdry}
 also show that most (i.e. $\sim 2/3$) of
low mass ellipticals and RQGs ($\rm M_{star} \lsim 10^{10.3}\; M_\sun$)
are not produced by mergers. Many authors have argued
that these galaxies are mostly quenched satellite galaxies whose gas halos
are stripped by much more massive central
galaxies \citep{vandenBosch2008, Peng2010}.

\subsection{Impacts of Major Mergers on Galaxy Assembly}

Mergers shift galaxies from lower mass bins to higher mass bins in
the GSMF. The efficiency of this process is clearly critical 
for the hierarchical structure formation paradigm.
\citet{Drory2008} tried to answer this question via comparisons 
between  observed GSMF variation against redshift and that 
predicted by the SFR vs. z relation which was well established (in particular
for $\rm z \leq 1$), attributing the difference to the merger
effects (including both major and minor mergers).
Our results on major-merger rates provide a new and more direct approach.

We define the following parameter $\rm \hat{\phi}_{merger}$
to evaluate the impact of major mergers on the GSMF $\rm \phi$:
\begin{equation}
\rm \hat{\phi}_{merger}(M_{star}) =  R_V(M_{star})/\phi(M_{star}) - R_{mg}(M_{star})
 \label{eq:phi},
\end{equation}
where $\rm \phi(M_{star})$ is the GSMF (D09), $\rm R_V(M_{star})$ the 
volume merger rate defined in Eq.~\ref{eq:R_V}, and $\rm  R_{mg}$ the 
differential merger rate estimated using Eq.~\ref{eq:Rmg}.

In Fig.~\ref{fig:gsmf} we compare our results with those of \citet{Drory2008}.
The solid line is our
result on the mean $\rm \hat{\phi}_{merger}(M_{star})$ over $\rm 0.3 < z < 0.9$.
In this redshift range, 
major mergers (as opposed to star formation) have significant impact to the
galaxy mass assembly only for the most massive galaxies with
$\rm M_{star} \gsim 10^{11.3}$. The GSMF change
due to major mergers dominates that due to
star formation only at $\rm M_{star} \gsim 10^{11.3}$.
For less massive galaxies with 
$\rm M_{star} < 10^{11.2}$ the GSMF change due to major mergers
is negligible (amplitude
$\leq 10$ \% Gyr$^{-1}$) in comparison to that due to star formation.
For these galaxies, 
our result is much flatter than that of \citet{Drory2008} for
the GSMF change rate due to mergers at both z=0.5 and z=1.
For massive galaxies ($\rm M_{star} \gsim 10^{11.3}$)
we find much steeper mass dependence than \citet{Drory2008}.
The major reason of the discrepancy is due to the difference between
the redshift dependent GMSFs used in this work \citep{Drory2009} and
those in \citet{Drory2008}, the latter were derived using data from
earlier FDF/GOODS surveys. 

\section{Major Mergers and (U)LIRGs}

\citet{Kartaltepe2010} carried out a study of luminous IR galaxies
(LIRGs: $\rm 10^{11} < L_{IR}/L_\sun < 10^{12}$) and ultra-luminous IR
galaxies (ULIRGs: $\rm L_{IR}/L_\sun > 10^{12}$) in the S-COSMOS
survey \citep{Sanders2007}. We neglect ULIRGs of $\rm L_{IR}/L_\sun >
10^{12.5}$ since very few galaxies have such
  high IR luminosities, and the AGN fraction increases rapidly with
  the $L_{IR}$ among these galaxies \citep{Kartaltepe2010}.
Taking galaxies with $\rm 10^{11.5} <
L_{IR}/L_\sun < 10^{12.5}$ from their
sample and adopting a (U)LIRG time scale of 140 Myr
\citep{Kartaltepe2010}, we estimate the (U)LIRG rates in different
stellar mass bins and in the redshift range of $\rm 0.2 < z < 1.0$.
The results are presented in Fig.~\ref{fig:ulirg}, compared 
with the average mass dependent volume merger rate. 

The mass dependence of (U)LIRG rates in all photo-z bins 
have the shape of the log-normal function, peaking at a rather
constant mass of $\sim \log(M^*_{star}/M_\sun) = 10.8$. 
In the low photo-z bin ($\rm 0.2\leq z \leq 0.4$)
(U)LIRGs are less frequent, consistent
with the fact that (U)LIRGs are very rare in the local universe
\citep{Sanders1996}.
The average mass dependent volume merger rate is above or
comparable to the (U)LIRG rates in all redshift and mass bins,
and therefore it is possible that most of the (U)LIRGs in this
redshift range are major-mergers, just like their local counterparts
\citep{Sanders1996}. Using morphological classifications, 
\citet{Kartaltepe2010} found that $\gsim 50\%$ of their (U)LIRGs
are major-mergers. This means that the merger-induced
(U)LIRG rates are even more below the average merger rate,
in particular for $\rm M_{star} \lsim 10^{10.3}$
and $\rm M_{star} \gsim 10^{11.3}$, than being depicted in Fig.~\ref{fig:ulirg}.
Hence, it is likely that a large fraction
of major-mergers, in particular those with $\rm M_{star} \lsim 10^{10.3}$
or $\rm M_{star} \gsim 10^{11.3}$, may not become (U)LIRGs.
It is interesting to note that the most massive mergers 
of $\rm M_{star} \gsim 10^{11.3}\; M_\sun$ have rather low (U)LIRG
rate. Most of them are wet or mixed mergers, but many
probably have relatively low gas content.
Galaxies of lower mass ($\rm M_{star} \sim 10^{10}\; M_\sun$) also have low 
(U)LIRG rate because their gas mass is not adequate to sustain the
very high SFR of the extreme starbursts in (U)LIRGs.

\section{Summary}

We have presented results of a statistical study on the cosmic
evolution of the mass dependent major-merger rate since $\rm z = 1$. A
stellar mass limited sample of major-merger pairs (the CPAIR sample)
was selected from the archive of the COSMOS survey.  It includes 617
galaxies in pairs/multiple-systems with stellar mass ratios $\rm \mu
leq 2.5$, projected separations in the range of $\rm 5 \leq r_{proj}
\leq 20\; h^{-1}\; kpc$, and in the photo-z range of $\rm 0.2 \leq
z_{phot} \leq 1.0$.  The pair selection was based on photo-z, with the
criterion of $\Delta z_{phot}/(1+z_{phot}) \leq 0.03$, and on visual
inspections of the HST-ACS images. The CPAIR sample is divided into
four photo-z bins of [$\rm 0.2\leq z \leq 0.4$, $\rm 0.4< z \leq 0.6$,
  $\rm 0.6< z \leq 0.8$, $\rm 0.8 < z \leq 1$].  Various biases in the
sample selection that caused incompleteness and spurious pair
contaminations have been studied. This resulted in a completeness
correction factor of [$0.78\pm 0.05$, $0.74\pm 0.05$, $0.71\pm 0.05$,
  $0.60\pm 0.05$] and a reliability correction factor of [$\rm 0.80\pm
  0.06$, $\rm 0.79\pm 0.06$, $\rm 0.77\pm 0.06$, $\rm 0.78\pm 0.06$],
respectively, for pairs in the four photo-z bins. The CPAIR sample is
complemented by a local (z=0) major-merger pair sample, selected in
the K-band from cross matches between 2MASS and SDSS-DR5 (an updated
version of the KPAIR sample in \citealt{Domingue2009}).

Mass dependent pair fractions at different redshifts derived using
CPAIR and KPAIR samples show no significant variations with stellar
mass. The integral pair fraction (i.e. pair fraction of all galaxies
regardless of stellar mass) demonstrates a moderately strong cosmic evolution,
with the best-fitting function of $\rm f_{pair} = 10^{-1.88(\pm 0.03)}
(1+z)^{2.2(\pm 0.2)}$. 

The merger time scale was taken from the simulation
results of \citet{Lotz2010}: $\rm T_{mg}/Gyr
= 0.3\times (M_{star}/10^{10.7} M_{\sun})^{-0.3}(1+z/8)$.  The best-fitting
function for the differential merger rate (for $\rm \mu \leq 3$ mergers)
is $\rm R_{mg}/Gyr^{-1} =
0.0453\times (M_{star}/10^{10.7} M_{\sun})^{0.3} (1+z)^{2.2}/(1+z/8)$. Accordingly, 
on average, galaxies of $\rm M_{star} \sim 10^{10\hbox{--} 11.5} M_{\sun}$ 
have undergone $\sim 0.5$ -- 1.5 times major mergers since z=1. 
Our result on the differential merger rate is in very good agreement
with that estimated using morphologically selected major-mergers 
\citep{Conselice2009} and
with the prediction of the ``semi-empirical'' model of \citet{Hopkins2010a}.

The mass dependent major-merger rates derived in this work indicate
that, for massive galaxies ($\rm M_{star} \geq
10^{10.5}\; M_\sun$) at $z \leq 1$, major mergers involving star forming galaxies
(i.e. wet and mixed mergers) can fully account for the formation rates of
both ellipticals and red quiescent galaxies (RQGs), lending support to
models that link both bulge formation and SFR quenching to major
mergers \citep[e.g.][]{Hopkins2008}. On the other hand, major mergers
cannot be responsible for the formation of most low mass ellipticals and
RQGs of $\rm M_{star} \lsim 10^{10.3}\; M_\sun$.  Dry mergers contribute 
negligibly to the major-merger rate in all mass and photo-z bins.
Major mergers have significant impact to the stellar mass assembly 
of the most massive galaxies ($\rm M_{star} \geq 10^{11.3}\; M_\sun$).
For less massive galaxies the stellar mass assembly is dominated by the star 
formation.

Comparisons with mass dependent (U)LIRG rates in different redshift
bins suggest that the frequency of major-merger events is comparable or
higher than that of (U)LIRGs. Most low mass mergers ($\rm M_{star}
\lsim 10^{10.3} M_\sun$) and most very massive mergers ($\rm M_{star} \gsim
10^{11.3} M_\sun$) may not become (U)LIRGs.

\noindent{\it Acknowledgments}:
This work is based on observations with the NASA/ESA {\em Hubble Space
Telescope}, obtained at the Space Telescope Science Institute, which
is operated by AURA Inc, under NASA contract NAS 5-26555; and Spitzer
Space Telescope, which is operated by the Jet Propulsion Laboratory,
California Institute of Technology under NASA contract 1407; also
based on data collected at : the Subaru Telescope, which is operated
by the National Astronomical Observatory of Japan; the XMM-Newton, an
ESA science mission with instruments and contributions directly funded
by ESA Member States and NASA; the European Southern Observatory under
Large Program 175.A-0839, Chile; Kitt Peak National Observatory, Cerro
Tololo Inter-American Observatory, and the National Optical Astronomy
Observatory, which are operated by the Association of Universities for
Research in Astronomy, Inc.  (AURA) under cooperative agreement with
the National Science Foundation; the National Radio Astronomy
Observatory which is a facility of the National Science Foundation
operated under cooperative agreement by Associated Universities, Inc ;
and the Canada-France-Hawaii Telescope with MegaPrime/MegaCam operated
as a joint project by the CFHT Corporation, CEA/DAPNIA, the NRC and
CADC of Canada, the CNRS of France, TERAPIX and the Univ. of Hawaii.
C.K.X acknowledges Kevin Bundy for constructive discussions and Alexie
Leauthaud for helps in analyzing the COSMOS HST-ACS lensing catalog.
Zara Scoville is thanked for proofing the English of the manuscript.
Y.Z. and Y.G. are grateful for the financial support from the NSF of China
(grants 10833006 and 10903029). Y.Z. thanks IPAC for the hospitality and
the financial support during his visit.


\bibliographystyle{apj}
\bibliography{/Volumes/Seagate/data1/bibliography/ckxu_biblio}

\appendix
\section{Incompleteness due to Missing Very Close Pairs  --- Analysis}

\renewcommand{\thefigure}{A-\arabic{figure}}
\setcounter{figure}{0}
\begin{figure}[!htb]
\epsscale{0.5}
\plotone{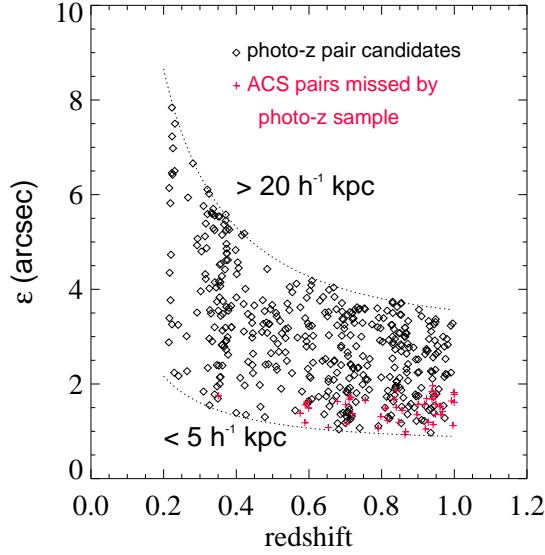}
\caption{Plot of the angular separation $\epsilon$ 
vs. redshift relation for pair candidates (open diamonds),
and for very close ACS pairs that are missed by the photo-z pair candidate
sample (plus symbols).
}
\label{fig:esp}
\end{figure}

In order to estimate how many pairs with
$\epsilon \lsim 2''$ are missing in our sample, we carried
out an analysis exploiting the COSMOS HST-ACS lensing catalog
\citep{Leauthaud2007, Leauthaud2010}. It
includes $\rm 1.2 \times 10^6$ galaxies detected by HST-ACS in the
F814 band (hereafter $\rm i_{ACS}$ band), with an angular resolution of
$\rm FWHM = 0.12''$ \citep{Leauthaud2007}.
From this catalog, we selected a sample of
``very close ACS pairs'' through the following procedure:
\begin{description}
\item{(1)} Find the match in the HST-ACS lensing
catalog for every D09 galaxy 
of $\rm \log(M_{star}) \geq \log(M_{lim})$ (see Section 2 for the definition
of $\rm M_{lim}$) with a searching radius of $\rm 0.8''$
and the criterion of $|i_{\rm ACS} - i_{\rm AB, D09}| \leq 1$ mag.
\item{(2)} Around the ACS matches of D09 galaxies, we search for 
ACS pairs with three criteria: (i) $|\Delta i_{\rm ACS}| 
\leq 1$ mag; (ii) $\rm \epsilon < 2.0''$;
(iii) $\rm r_{proj} \geq 5 \; h^{-1}\; kpc$.
\end{description}
\begin{figure}[!htb]
\epsscale{0.5}
\plotone{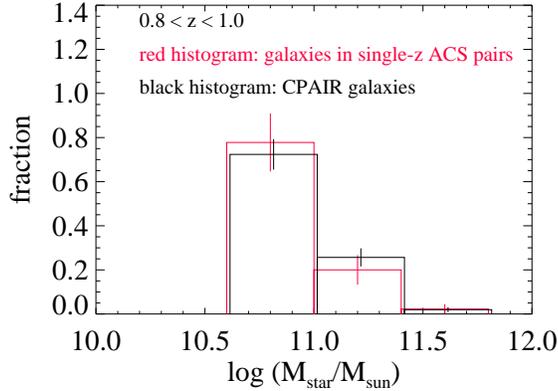}
\caption{Comparison of mass distributions of CPAIR galaxies and of
galaxies in the single photo-z ACS pairs ($\epsilon < 2''$)
with $\rm 0.8 < z_{phot} < 1.0$, including 31 of the
47 ``very close'' ACS pairs.}
\label{fig:macs}
\end{figure}

The procedure selected 222 very close ACS pairs. Among them, 171 pairs
have both component galaxies with photo-z matches, including 53 pairs found in
the photo-z pair sample and the remaining 118 pairs consisting of
galaxies of discordant photo-z's or with mass ratios larger than 2.5.
In the remaining 51 pairs, 4 are multi-peak single galaxies in
the ACS images. The final sample has 47 pairs in which only one of two
galaxies was detected in the photo-z catalog.  These pairs, shown in
Fig.~\ref{fig:esp}
 by red crosses, have the average angular separation $\rm <\epsilon>
= 1.45''$ with the standard deviation of $\rm \sigma = 0.27''$.
Fig.~\ref{fig:esp} also shows that, for pairs of $z > 0.5$,
the lower boundary for the pair separations,
$\rm r_{proj} \geq 5 \; h^{-1}\; kpc$, corresponds to a angular
separation of $\rm \epsilon \sim 1''$.

We then used Monte Carlo simulations to estimate the expected number of
spurious pairs in the sample of very close ACS pairs, utilizing the
138001 galaxies in D09 sample.  In each of the 100 simulations, we
randomly put these galaxies in a 1.7 deg$^2$ region, with all other
properties of the galaxies, including the photo-z and stellar mass,
intact.  We then search companions around each of the galaxies
of $\rm \log(M_{star}) \geq \log(M_{lim})$ in the
simulated sample according to the following criteria: $|\Delta
i_{\rm AB}| \leq 1$ mag, (ii) $\rm \epsilon < 2.0''$, and (iii) $\rm
r_{proj} \geq 5 \; h^{-1}\; kpc$. Spurious pairs that do not pass the four
pair selection criteria in Section 2 were then counted. These
simulations found a mean spurious pair number of 103.3 with a 1-$\sigma$
dispersion of 7.0. As described above, the number of confirmed spurious
pairs of photo-z galaxies in the sample of very close ACS pairs is
118.  This number is slightly higher than the mean total number of
spurious pairs (103.3$\pm 7.0$) predicted by the Monte Carlo
simulations, perhaps due to galaxy
clustering. Thus majority of 47 single photo-z ACS
pairs are in fact real and not chance superpositions. 
Indeed their ACS images very often
show signs of interaction. In what follows we shall make the conservative
assumption that all 47 ACS pairs are real major-merger 
pairs.

There are $\rm N_{ACS} = [1, 5, 10, 31]$ 
single photo-z ACS pairs in the four photo-z bins.
The corresponding
incompleteness due to missing of very close pairs, estimated
according to the ratio $\rm N_{ACS}/(N_{photo-z}+N_{ACS})$ ($\rm N_{photo-z}$
being the number of photo-z pair 
candidates), is 0.01$\pm 0.01$, 0.06$\pm 0.03$,
0.08$\pm 0.03$, and 0.20$\pm 0.04$ for the four photo-z bins, respectively.
As a check, we found no significant difference between the stellar mass
distributions of galaxies in the ACS single photo-z pairs and of those
in the CPAIR sample in the redshift bin of $\rm 0.8 < z_{phot} < 1.0$ 
(Fig.~\ref{fig:macs}).

\section{Incompleteness due to Photo-z Errors
and Spurious Pairs due to Projection --- Monte Carlo Simulations}
\renewcommand{\thefigure}{B-\arabic{figure}}
\setcounter{figure}{0}
\begin{figure}[!htb]
\epsscale{0.5}
\plotone{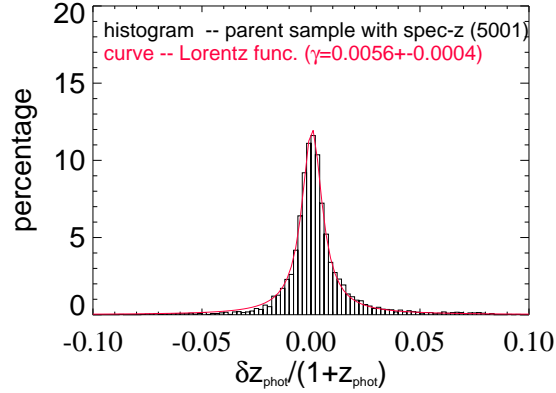}
\caption{Plot of 
$\rm (z_{phot} - z_{spec})/(1+z_{spec})$ distribution of 
5001 galaxies in the parent sample that have spec-z.
The red solid curve is the best-fit Lorentzian function. 
}
\label{fig:dzdist}
\end{figure}
\begin{figure}[!htb]
\epsscale{0.8}
\plotone{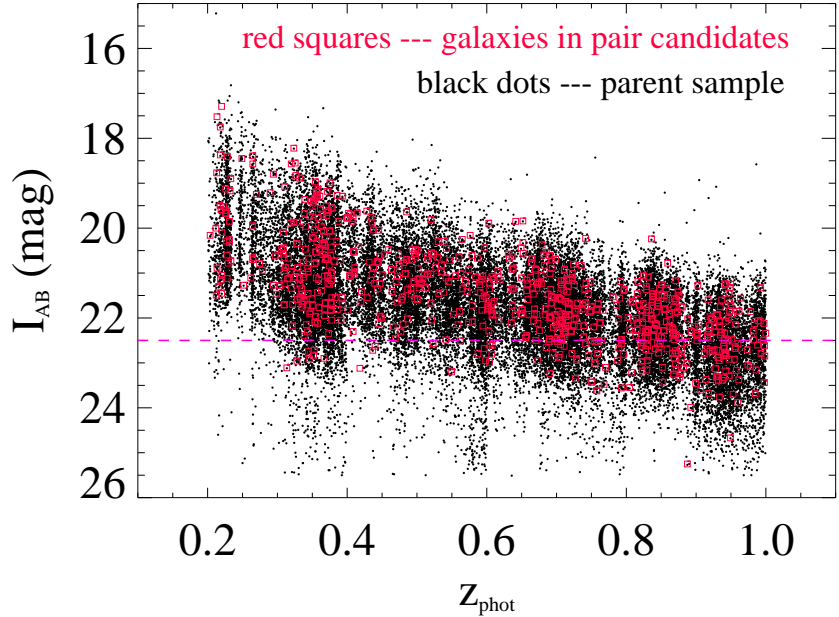}
\caption{The i-band magnitude ($i_{\rm AB}$) vs. photo-z 
plot. Galaxies in the pair candidates are shown by red squares, and
galaxies in the parent sample (Table~\ref{tbl:parent}) 
by black dots. The dashed line marks 
the boundary of $i_{\rm AB} = 22.5$.
}
\label{fig:idist}
\end{figure}

We examine the accuracy of the photo-z using the
spectroscopically measured redshift (hereafter spec-z)
of galaxies in the parent sample that were observed in
the zCOSMOS survey \citep{Lilly2007}. zCOSMOS includes a
magnitude-limited survey (zCOSMOS-bright)
for about 20,000 galaxies of $i_{\rm AB} < 22.5$ and $\rm 0.1 < z < 1.2$, 
covering 1.7 deg$^2$ COSMOS
includes 10643 galaxies. There
are 5001 matches (matching radius $= 2"$) between galaxies in the parent
sample (Table~\ref{tbl:parent}) 
and zCOSMOS sources with reliable spec-z measurements (z-class
indices being 4's, 3's, 9.5, 9.4, 9.3, 2.5, 2.4 or 1.5).  This is
19.5\% of 25711 galaxies in the parent sample that have $i_{\rm AB} < 22.5$.
Fig.~\ref{fig:dzdist} shows $\rm (z_{phot} - z_{spec})/(1+z_{spec})$
of the 5001 galaxies. 
The distribution, an estimate of the photo-z
error distribution (the spec-z error 
$\rm \lsim 100\; km\; sec^{-1}$, \citealt{Lilly2007}),
has a $\rm 1.48\times median(|z_{phot} - z_{spec}|/(1+z_{spec})) = 0.007$, 
consistent with the result of \citet{Ilbert2009}. 
It can be best fitted by a Lorentzian function of the form
\begin{equation}
\rm L(x) = {1\over \gamma \pi} {1\over 1 + ((x-x_0)/\gamma)^2},
\end{equation}
with $\rm x=(z_{phot}-z_{spec})/(1+z_{spec})$, $\gamma = 0.0056\pm
0.0004$ and $\rm x_0 = 0.0005\pm 0.0003$.  

In Fig.~\ref{fig:idist}, the i-band magnitudes are shown for
galaxies in pair candidates (red squares)
and in the parent sample (black dots). Both
the pair sample and the parent sample are dominated by galaxies brighter
than $i_{\rm AB} = 22.5$. In the pair candidates,
factions of 0.032, 0.061, 0.168, 0.352 of the sample
are fainter than $i_{\rm AB} = 22.5$ for the four redshift bins.

The Monte Carlo simulations for the estimation of completeness
correction factor due to photo-z errors were
based on the above results. The simulations included
a total of 1000 repeats for each photo-z
bin, each containing 100 pairs.  For pairs in a given photo-z bin,
each galaxy had a chance
to be fainter (or brighter) than $i_{\rm AB} = 22.5$, the probability 
(simulated by a random number generator) being
equal to the observed fraction of such galaxies in the photo-z bin.
For galaxies brighter than $i_{\rm AB} = 22.5$, an error of $\rm
z_{phot}/(1+z_{phot})$ was assigned to it using a random number
generator with weighted probability distribution function given by a
Lorentzian with $\rm x_0=0$ and $\gamma = 0.0056$. For
galaxies fainter than $i_{\rm AB} = 22.5$, the random error was
generated according to a Lorentzian function with $x_0=0$ and $\gamma
= 0.0096$ ($\simeq 0.0056\times 0.012/0.007$, where $\rm 0.012/0.007$
is the ratio between the photo-z accuracies for galaxies of $22.5 \leq
i_{\rm AB} < 24$ and of $i_{\rm AB}< 22.5$, \citealt{Ilbert2009}). We
ignored the real velocity difference and assumed that the photo-z
difference in a pair is purely due to photo-z errors. The completeness
factor was estimated by the fraction of simulated pairs with
$\rm |\Delta z_{phot}/(1+z_{phot})| \leq 0.03$. According to
the simulations, the completeness correction factor is [$\rm 0.78\pm
  0.05,$ $\rm 0.77\pm 0.06$, $\rm 0.76\pm 0.05$, $\rm 0.73\pm 0.05$] in
the four photo-z bins.

Another set of Monte Carlo simulations (each consisting of 1000
repeats) were carried out to estimate the SPF due to projection,
utilizing the 138001 galaxies in D09 sample. Here we pretended that
the photo-z's in that sample are 100\% accurate, and then added errors
to them using the same algorithm as described above.  The sky
coordinates of the galaxies were also randomized, filling a 1.7
deg$^2$ region uniformally. The other properties of the galaxies,
including the stellar mass, were left intact.  We then selected pairs
from this simulated parent sample by applying the four selection
criteria presented in Section 2.  Spurious pairs were identified when
the ``true velocity difference'', calculated using the ``true
redshifts'' (i.e. the photo-z's without added error), is $\rm > 500\;
km\; sec^{-1}$. In the four photo-z bins, the simulations found [$7.5
  \pm 2.8$, $6.2 \pm 2.5$, $11.4 \pm 3.3$, $11.1 \pm 3.4$] spurious
pairs. Dividing these numbers by the numbers of photo-z pair
candidates, the predicted SPF in the four photo-z bins are $0.07\pm
0.03$, $0.08\pm 0.03$, $0.10\pm 0.03$ and $0.09\pm 0.03$,
respectively.

\section{Distribution of Line-of-sight Velocity Difference of Major-Merger Pairs}

\renewcommand{\thefigure}{C-\arabic{figure}}
\setcounter{figure}{0}
\begin{figure}[!htb]
\epsscale{0.5}
\plotone{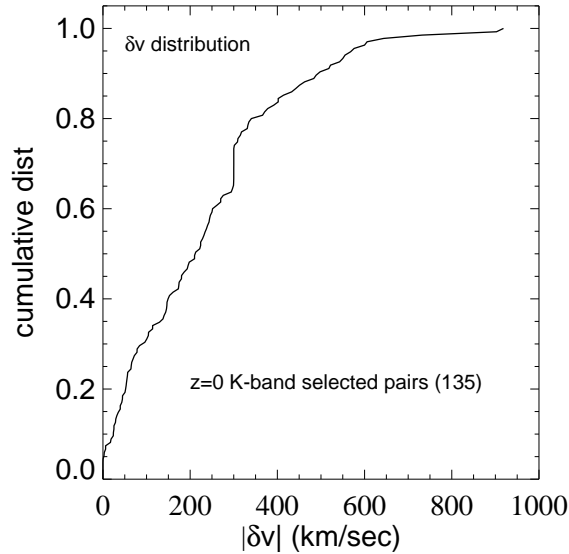}
\caption{Cumulative distribution of $\rm \Delta v$ of major-merger pairs, derived
using a sample of z=0 K-band selected pairs \citep{Domingue2009}.
}
\label{fig:vdist}
\end{figure}

Local major-merger pairs in \citet{Domingue2009} were selected using
nearly identical selection criteria as the CPAIR sample (cf. Section
4) except for that spec-z were used and they include all pairs
(isolated or in groups/clusters) with $\rm \Delta v_{spec} \leq 1000$
km sec$^{-1}$. Here we exploit these data to determine the cumulative
distribution of $\rm \Delta v$ and, in particular, the fraction of
pairs with $\rm 500 < \Delta v \leq 1000\; km\; sec^{-1}$.  We assume
that all pairs with $\rm \Delta v > 1000\; km\; sec^{-1}$ are spurious.
In the sample of \citet{Domingue2009}, which is $0.82\%$ complete for
the spec-z, 135 pairs have measured
$\rm v_{spec}$ for both components and $\rm v_{spec} > 2000\; km\; sec^{-1}$
(excluding pairs in the local super-cluster). The cumulative distribution of
$\rm \Delta v$ of these pairs is plotted in Fig.~\ref{fig:vdist}. 
From this distribution,
we found that the fraction of pairs with 
$\rm \Delta v > 500$ km sec$^{-1}$ is $\rm 9.6\%$ with a random
error of $\rm 2.5\%$ (binomial statistics). 
It should be pointed out that the $\rm \Delta v$
distribution is sensitive to the environment and to the pair separation 
\citep{Ellison2010}. Therefore caution should be taken when applying
the result here to other pair samples.

\section{Incompleteness and Spurious Pairs Fraction ---
A Comparison with Spec-z Pairs}

\renewcommand{\thefigure}{D-\arabic{figure}}

We made a comparison between pairs in the CPAIR sample 
and a sample of spec-z pairs (ZPAIR sample) selected from 
the zCOSMOS survey \citep{Lilly2007}. In principle, this
comparison applies only to galaxies brighter than $i_{\rm AB} = 22.5$,
the magnitude limit of zCOSMOS. However, most of galaxies in
photo-z pairs are brighter than $i_{\rm AB} = 22.5$ (Fig.~\ref{fig:idist}). 
Images of the 16 pairs in the ZPAIR sample, listed in Table~\ref{tbl:zpair},
are shown in Fig.~\ref{fig:zpair}. 
Among them are all CPAIRs (14) where both components have 
spec-z's. They comprise a very small fraction
of total CPAIR sample because only 19.5\% of galaxies
of  $i_{\rm AB} < 22.5$ in the parent sample have spec-z. 
All but two of the 14 pairs have the velocity difference 
$\rm \Delta v_{spec} < 500\; km\; sec^{-1}$.
This results in an empirical estimate for the SPF of
$2/14 = 0.14$, or a reliability of $\rm 0.86$, 
with a binomial uncertainty of $\pm 0.12$.

Table~\ref{tbl:zpair} 
also includes two pairs (ZPAIR-03 and ZPAIR-07) that were missed 
by CPAIR sample, both having $\rm \Delta z_{phot}/(1+z_{phot})> 0.03$ but
$\rm \Delta v_{spec} < 500\; km\; sec^{-1}$. As shown in Fig.~\ref{fig:zpair},
the primary of ZPAIR-03 is a close pair itself, and its
large $\rm z_{phot}$ error (Table~\ref{tbl:zpair}) is likely due to confusion
in the photometric data. ZPAIR-07, having 
$\rm \Delta z_{phot}/(1+z_{phot})= 0.031$, barely missed the cut for the
photo-z pairs. Adding the 12 genuine CPAIRs, there are 14 pairs 
with $\rm \Delta v < 500\; km\; sec^{-1}$ in 
the ZPAIR sample. This results in a completeness for the
CPAIR sample of $12/14 = 0.86$, with a statistical error 0.12.

\section{Distribution of Primary-to-Secondary Mass Ratio of Close Pairs}
\renewcommand{\thefigure}{E-\arabic{figure}}
\setcounter{figure}{0}
\begin{figure}[!htb]
\plottwo{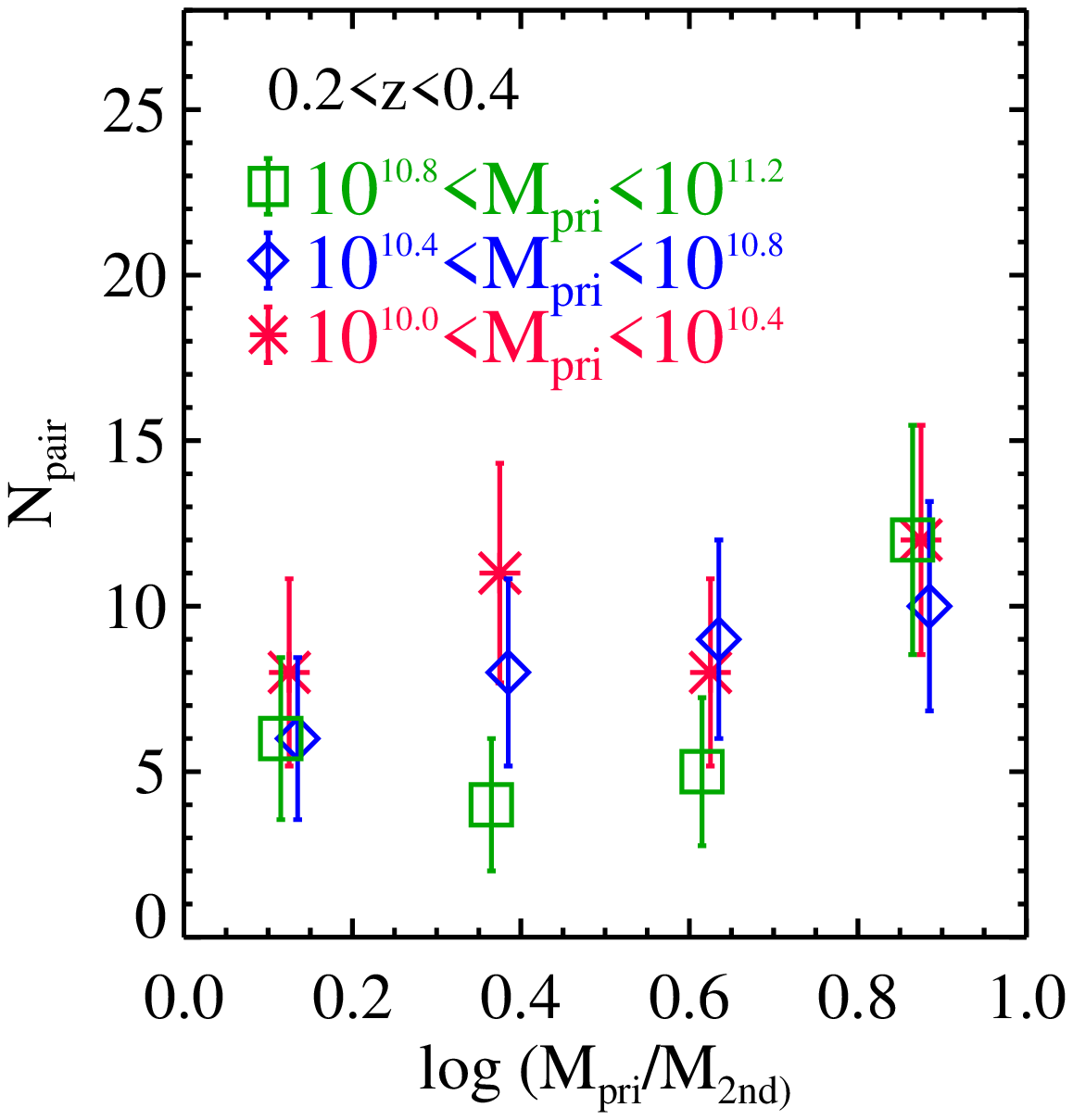}{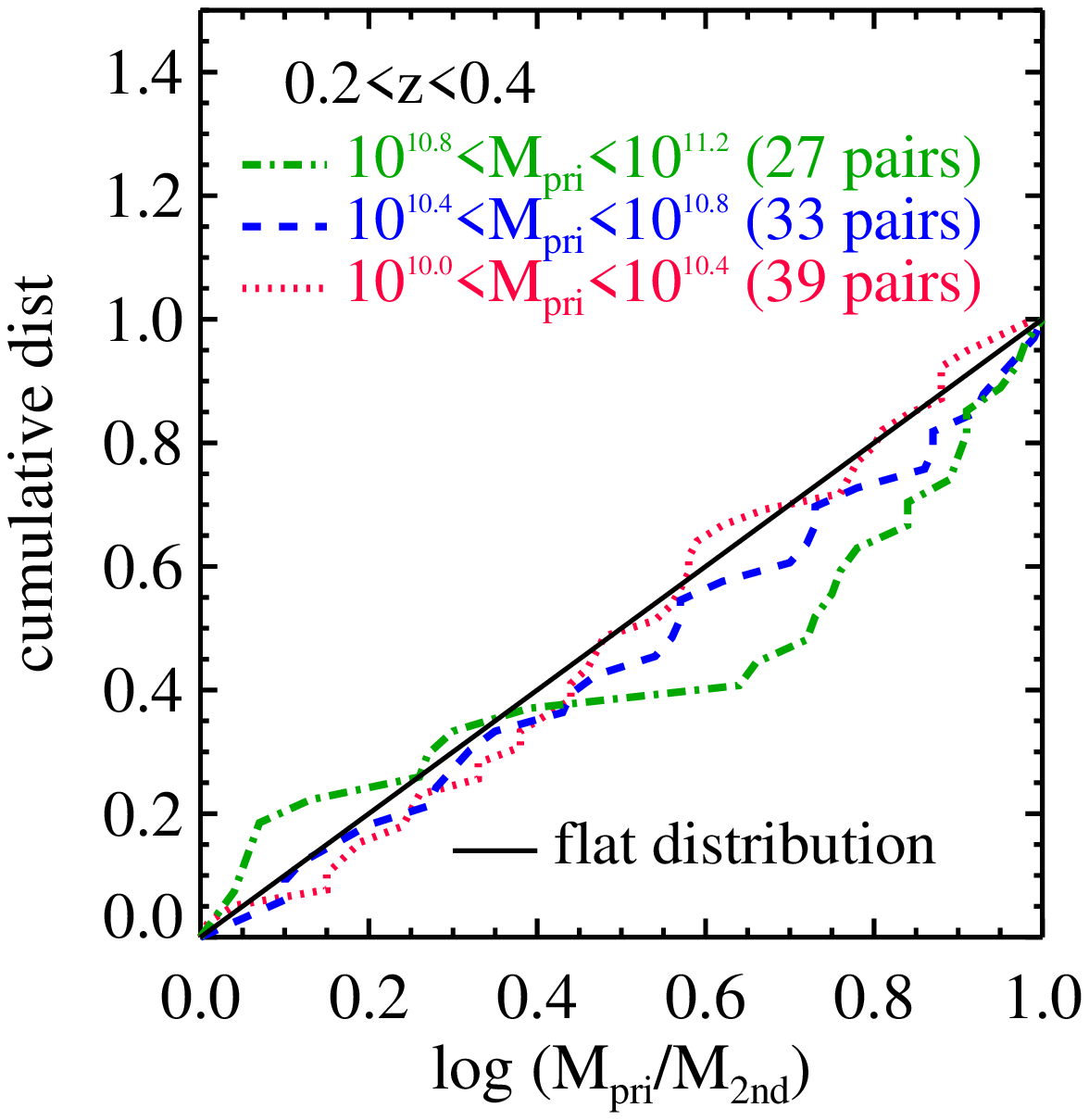}
\caption{
{\bf Left:} Differential distributions of $\rm \log(\mu)$ 
($\rm \mu = M_{pri}/M_{2nd}$) of close pairs
($\rm 5 \leq r_{proj} \leq 20\; h^{-1}\; kpc$) in three mass bins. Derived
using pair candidates selected in a volume limited sample of
galaxies with redshift in the range of $\rm 0.2 \leq z \leq 0.4$.
{\bf Right:}
Normalized cumulative distributions of $\rm \log(\mu)$ 
($\rm \mu = M_{pri}/M_{2nd}$) of the same close pairs.
}
\label{fig:mratdist}
\end{figure}
In the parent sample of CPAIR, galaxies in the photo-z
bin of $\rm 0.2 \leq z_{phot} \leq 0.4$
form a volume-limited sample above the stellar mass limit 
$\rm M_{star} = 10^{9.0} M_\sun$. Using this volume limited sample and
applying the same pair selection criteria in Section 2
except for expanding the mass ratio limit to $\rm \log(\mu_{max}) = 1.0$,
where $\mu = M_{pri}/M_{2nd}$ is the mass ratio,
we selected 
pair candidates (including both major and minor pairs)
in three mass bins: $\rm 10.0 < \log(M_{pri}/M_{\sun}) \leq 10.4$, 
$\rm 10.4 < \log(M_{pri}/M_{\sun}) \leq 10.8$, and
$\rm 10.8 < \log(M_{pri}/M_{\sun}) \leq 11.2$. Assuming that completeness
and reliability corrections for these pair candidates do not depend on the
mass ratio $\mu$, we calculated the differential and cumulative 
$\rm \log(\mu)$ distributions of these close pairs ($\rm 5 \leq r_{proj} \leq
20\; h^{-1}\; kpc$). The results are plotted in Fig.~\ref{fig:mratdist}. 
It shows that the flat 
distribution, i.e. $\rm df_{pair}/d\log(\mu) = constant$, 
is a reasonably good approximation. It is worth noting that our
result is different from that of \citet{Ellison2010}, who found a mass ratio
distribution for SDSS pairs that is tilted toward low $\mu$ pairs
(i.e. major mergers). However, their result is affected significantly
by the ``missing secondary'' bias, causing
severe incompleteness of the $\mu \sim 10$ minor-mergers in their sample.
                                                                               
\renewcommand{\thefigure}{D-\arabic{figure}}
\setcounter{figure}{0}
\begin{figure*}
\vbox{
  \begin{minipage}[l]{1.4\textwidth}
    \vskip-4truecm \hspace{-7truecm}
    \plotone{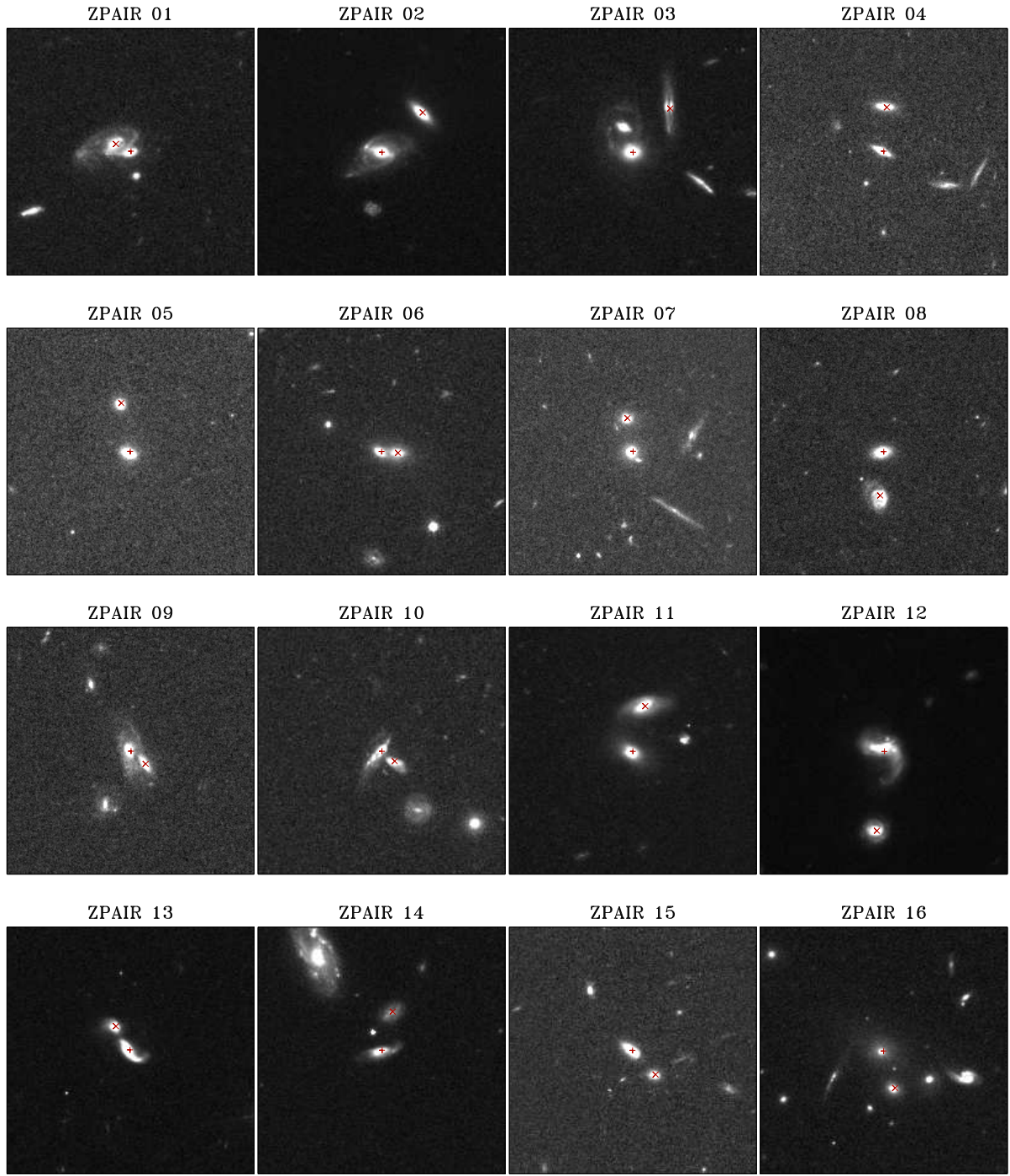}
 \end{minipage} \  \hfill \
     }
  \vskip-3truecm
  \caption{
    HST-ACS (F814) images of close major-merger pairs in
    zCOSMOS survey (ZPAIRs in Table~\ref{tbl:zpair}). The size of 
    all images is $20''\times 20''$. The crosses mark the positions of 
    component galaxies in the pairs. {\it Notes}: ZPAIR-03 and ZPAIR-07 are
    not in the CPAIR (photo-z pairs) sample. The large neighbor 
    in ZPAIR-14 is
    likely a foreground galaxy. ZPAIR-16 is in a compact 
    group of galaxies
    (three in this group were included in the CPAIR sample).}
\label{fig:zpair}
\end{figure*}

\renewcommand{\thetable}{D-\arabic{table}}
\setcounter{table}{0}
\begin{deluxetable*}{ccccccccccccccc}
\tabletypesize{\normalsize}
\setlength{\tabcolsep}{0.05in} 
\tablewidth{0pt}
\tablecaption{Pairs in Z-COSMOS Survey (ZPAIR) \label{tbl:zpair}}
\tablehead{
 \colhead{ZPAIR} 
& &  \colhead{$\rm \Delta v_{spec}$} 
& &  \colhead{RA$_1$} 
& &  \colhead{Dec$_1$} 
& &  \colhead{$\rm z_{spec,1}$} 
& &  \colhead{$\rm zclass_{1}$} 
& &  \colhead{$\rm z_{phot,1}$} 
& &  \colhead{$\rm \log(M_{star,1})$} 
\\
\colhead{ID} 
& &  \colhead{($\rm km\; sec^{-1}$)} 
& &  \colhead{(degree)} 
& &  \colhead{(degree)} 
& &
& &
& &
& &  \colhead{($\rm M_\sun$)} 
\\
\\
\hline
& &  \colhead{$\rm \Delta z_{phot}/(1+z_{phot})$} 
& &  \colhead{RA$_2$} 
& &  \colhead{Dec$_2$} 
& &  \colhead{$\rm z_{spec,2}$} 
& &  \colhead{$\rm zclass_{2}$} 
& &  \colhead{$\rm z_{phot,2}$} 
& &  \colhead{$\rm \log(M_{star,2})$} 
\\
& & 
& &  \colhead{(degree)} 
& &  \colhead{(degree)} 
& &
& &
& &
& &  \colhead{($\rm M_\sun$)} 
}
\startdata
     01& &      43& &  149.713610& &  2.019610& &  0.6444& &   4.5& &  0.6518& &  10.92 \\
       & &   0.012& &  149.713950& &  2.019773& &  0.6441& &   4.5& &  0.6324& &  10.83 \\
\\
     02& &     218& &  149.839700& &  1.929228& &  0.3722& &   3.5& &  0.3856& &  10.80 \\
       & &   0.020& &  149.838780& &  1.930125& &  0.3711& &   4.5& &  0.3585& &  10.76 \\
\\
     03$^\dagger$& &      70& &  150.009050& &  2.274964& &  0.4726& &   2.5& &  0.5769& &  11.05 \\ 
       & &   0.071& &  150.008210& &  2.275954& &  0.4730& &   2.5& &  0.4645& &  10.78 \\
\\
     04$^*$& &     817& &  150.107580& &  2.556516& &  0.5038& &   3.5& &  0.4915& &  10.33 \\
       & &   0.004& &  150.107510& &  2.557509& &  0.4990& &   3.5& &  0.4969& &  10.21 \\
\\
     05& &      73& &  150.115480& &  1.975120& &  0.4385& &   4.5& &  0.4431& &  10.45 \\
       & &   0.025& &  150.115680& &  1.976216& &  0.4381& &   3.5& &  0.4069& &  10.05 \\
\\
     06& &      65& &  150.126060& &  1.913758& &  0.7360& &   2.5& &  0.7168& &  11.17 \\
       & &   0.003& &  150.125690& &  1.913726& &  0.7365& &   3.5& &  0.7108& &  11.15 \\
\\
    07$^\dagger$& &      11& &  150.168760& &  2.315481& &  0.8524& &   2.5& &  0.7921& &  10.97 \\
       & &   0.031& &  150.168890& &  2.316234& &  0.8523& &   1.5& &  0.8473& &  10.94 \\
\\
     08& &     220& &  150.196380& &  2.371582& &  0.6834& &   4.5& &  0.6783& &  10.86 \\
       & &   0.003& &  150.196460& &  2.370591& &  0.6850& &  22.5& &  0.6726& &  10.64 \\
\\
     09& &     103& &  150.230880& &  1.845002& &  0.6226& &   2.5& &  0.6072& &  10.67 \\
       & &   0.014& &  150.230550& &  1.844713& &  0.6233& &   3.5& &  0.5840& &  10.60 \\
\\
     10& &     118& &  150.258800& &  1.988773& &  0.7258& &   2.5& &  0.7168& &  10.70 \\
       & &   0.001& &  150.258510& &  1.988547& &  0.7267& &   2.5& &  0.7191& &  10.32 \\
\\
    11$^*$& &    6318& &  150.359560& &  2.659517& &  0.4309& &  22.5& &  0.3991& &  10.94 \\
       & &   0.002& &  150.359280& &  2.660543& &  0.3974& &   3.5& &  0.4018& &  10.81 \\
\\
     12& &     142& &  150.396810& &  2.519130& &  0.2189& &  24.5& &  0.2283& &  9.97 \\
       & &   0.003& &  150.396980& &  2.517332& &  0.2195& &   4.5& &  0.2323& &  9.93 \\
\\
     13& &     333& &  150.421160& &  2.654335& &  0.2144& &   4.5& &  0.2312& &  9.57 \\
       & &   0.003& &  150.421480& &  2.654870& &  0.2158& &   4.5& &  0.2271& &  9.21 \\
\\
     14& &     118& &  150.457210& &  2.695287& &  0.2189& &   4.5& &  0.2277& &  9.45 \\
       & &   0.008& &  150.456970& &  2.696160& &  0.2194& &   4.5& &  0.2184& &  9.07 \\
\\
     15& &     219& &  150.494640& &  2.187936& &  0.3695& &   4.5& &  0.3674& &  9.97 \\
       & &   0.007& &  150.494120& &  2.187388& &  0.3684& &   3.5& &  0.3772& &  9.58 \\
\\
     16& &     196& &  150.504990& &  2.225083& &  0.8374& &  22.5& &  0.8360& &  11.72 \\
       & &   0.002& &  150.504730& &  2.224246& &  0.8357& &   2.5& &  0.8315& &  11.34 \\
\hline
\enddata
\tablecomments{
\begin{description}
\item{$\dagger$} Missing in the CPAIR sample.
\item{$*$} Spurious pairs (with $\Delta v_{spec} > 500\; km\; sec^{-1}$) in the CPAIR sample.
\end{description}
}
\end{deluxetable*}


\end{document}